\documentclass[12pt,letterpaper]{article}

\usepackage{graphicx}
\usepackage[english]{babel}

\usepackage{amsmath,amssymb,bm}
\usepackage{dcolumn,multirow}
\usepackage[usenames]{color}

\numberwithin{equation}{section}

\newcommand{\hpi}{\hat{\pi}}
\newcommand{\hC}{\hat{C}}
\newcommand{\excv}{\delta}

\newcommand{\ra}{\rightarrow}
\newcommand{\tle}{\tilde{e}}
\newcommand{\phvw}{\hat{p}^{\{v,w\}}}
\newcommand{\ptvw}{\tilde{p}^{\{v,w\}}}

\newcommand{\jdiam}{\blacklozenge}
\newcommand{\jsqua}{\blacksquare}
\newcommand{\jtria}{\blacktriangle}
\newcommand{\jcirc}{\bullet}

\newcommand{\pr}{\mathrm{Pr}}
\newcommand{\eset}{[n]^{\{2\}}}
\newcommand{\pvw}{p^{\{v,w\}}}
\newcommand{\bp}{\mathbf{p}}
\newcommand{\bq}{\mathbf{q}}
\newcommand{\bB}{\mathbf{B}}
\newcommand{\bK}{\mathbf{K}}
\newcommand{\bQ}{\mathbf{Q}}
\newcommand{\bPhi}{\bm{\Phi}}
\newcommand{\cH}{\mathcal{H}}
\newcommand{\Kappa}{\mathrm{K}}
\newcommand{\uhead}[1]{\vspace{.2in}\noindent\textbf{#1}}
\newcommand{\expt}{\mathbb{E}}
\newcommand{\params}{\vec{\mu}}

\newcommand{\pvp}{\mathcal{P}}
\newcommand{\pcp}{\mathcal{P}}
\newcommand{\pvk}{\mathcal{Q}}
\newcommand{\pck}{\mathcal{Q}}
\newcommand{\pvkp}{\mathcal{R}}
\newcommand{\pckp}{\mathcal{R}}
\newcommand{\tmp}{\mathcal{A}}
\newcommand{\tcp}{\mathcal{A}}
\newcommand{\tmk}{\mathcal{B}}
\newcommand{\tck}{\mathcal{B}}
\newcommand{\tmkp}{\mathcal{C}}
\newcommand{\tckp}{\mathcal{C}}

\begin{document}

\date{}    

\title{\bf Community detection and tracking on networks from a data fusion perspective}

\author{
James P. Ferry \\
Metron, Inc.\\
Reston, VA, U.S.A.\\
Email: ferry@metsci.com
\and
J. Oren Bumgarner \\
Metron, Inc.\\
Reston, VA, U.S.A.\\
Email: bumgarner@metsci.com
}

\maketitle                        
\thispagestyle{empty}   

\selectlanguage{english}

\noindent
{\bf Abstract -
   {\small\em
Community structure in networks has been investigated from many
viewpoints, usually with the same end result: a community detection
algorithm of some kind.  Recent research offers methods for combining
the results of such algorithms into timelines of community evolution.
This paper investigates community detection and tracking from the data
fusion perspective.  We avoid the kind of hard calls made by
traditional community detection algorithms in favor of retaining as
much uncertainty information as possible.  This results in a method
for directly estimating the probabilities that pairs of nodes are in
the same community.  We demonstrate that this method is accurate using
the LFR testbed, that it is fast on a number of standard network
datasets, and that it is has a variety of uses that complement those
of standard, hard-call methods.  Retaining uncertainty information
allows us to develop a Bayesian filter for tracking communities.  We
derive equations for the full filter, and marginalize it to produce a
potentially practical version.  Finally, we discuss closures for the
marginalized filter and the work that remains to develop this into a
principled, efficient method for tracking time-evolving communities on
time-evolving networks.
}
}

\vspace{0.5cm}


\section{Introduction}

The science of networks has a large and multidisciplinary literature.
Freeman traces the sociological literature on networks from its
pre-cursors in the 1800s and earlier, through the sociometry of the
1930s and Milgram's ``Small Worlds'' experiments in the 1960s, to its
current form~\cite{Fre04}.  Sociologists and statisticians introduced
the idea of defining {\em network metrics}: simple computations that
one can perform on a network, accompanied by arguments that explain
their significance: e.g., the {\em clustering coefficient} and various
measures of network {\em centrality}~\cite{WaFa94}.  What Lewis calls
the ``modern period'' of network science~\cite{Lew09} began in 1998
with the influx of physicists into the field (e.g., Barab\'asi and
Newman).  The physicists brought novel interests and techniques (power
laws, Hamiltonians, mean field approximation, etc.), particularly from
statistical physics, along with an overarching drive toward {\em
universality}---properties of network structure independent of the
particular nature of the nodes and links involved~\cite{New10}.
Mathematicians have their own traditions of graph theory~\cite{Bol98},
and, in particular, random graph theory~\cite{JLR00,Bol01} which
emphasizes rigorously formulated models and what properties the graphs
they produce have (with high probability) in different asymptotic
regions of the models' parameter spaces.  Finally, computer scientists
have developed a wide variety of efficient network
algorithms~\cite{CLRS09}, and continue to contribute broadly because
ultimately the processing of data into usable results is always
accomplished via an algorithm of some kind, and because solid computer
science is needed for processing megascale, real-world networks.

Each of the above communities brings important, complementary talents
to network science.  The {\em data fusion} community has important
perspectives to offer too, due both to the broad range of practical
issues that it addresses, and to characteristics of the mathematical
techniques it employs~\cite{BWT11}.

The defining problem of data fusion is to process data into useful
knowledge.  These data may be of radically different types.  One might
consider a single data point to be, e.g., the position estimate of a
target, a database record containing entries for various fields, an
RDF triple in some specified ontology, or a document such as an image,
sound, or text file.  {\em Data mining} deals with similar issues, but
focuses on the patterns in and transformations of large data sets,
whereas data fusion focuses on distilling effective situational
awareness about the real world.  A central paradigm in data fusion is
the hierarchy of fusion {\em levels} needed to transform raw data into
high-level knowledge, the most widespread paradigm being the JDL
(Joint Directors of Laboratories) hierarchy~\cite{StBo09}.  In this
paradigm, level 0 comprises all the pre-processing that must occur
before one can even refer to ``objects.''  In some cases it is clear
that a data point corresponds to some single object, but it is unclear
which object: this is an {\em entity resolution}
problem~\cite{BhGe06}.  In other cases, a data point contains
information about multiple objects, and determining which information
corresponds to which object is a {\em data association}
problem~\cite{MoCh03,Fer10}.  Processing speech or images requires
solving a {\em segmentation} problem to map data to
objects~\cite{Gra06}, and natural language processing involves a
further array of specific techniques (Named Entity Recognition, etc.).
One benefit of a data-fusion approach to network science is its
careful consideration, at level 0, of how to abstract representations
from raw data.  In the network context, this applies not just to nodes
(i.e., objects), but to the links between them. Sometimes one imposes
an arbitrary threshold for link formation; sometimes multi-node
relationships (i.e., hypergraph edges) are replaced with edges between
all nodes involved.  Edges can have natural direction, weights, or
types (and nodes may have attributes) that are retained or ignored in
a graph representation.  When data are inappropriately shoehorned
into a network format, or important node or link attributes are
ignored, then the results derived from that graph representation may
be much less powerful than they could be, or even completely
misleading.

Level-0 data fusion encompasses these pre-processing techniques, drawn
from computer science, data mining, and the domains specific to the
data being considered.  Higher-level fusion (say, JDL levels 3--5)
addresses another set of issues important to a complete theory of
network science.  These issues relate to human knowledge and intent.
Just as level-0 fusion has similarities with the computer-science
approach to networks, higher-level fusion has some overlap with the
sociological approach.  Levels 1 and 2, on the other hand, correspond
loosely to the more theoretical approaches of mathematics and physics.
Level 1 addresses the detection, state estimation, and tracking of
individual objects~\cite{CWB11}; whereas level 2 broadens the scope to
tracking groups of objects~\cite{FFK11} and to the general assessment
of multiple-object situations~\cite{BKSK06}.  In data fusion, however,
the overriding problem is how to achieve cohesion between the various
levels~\cite{HNLL10}.  Achieving such cohesion would be a valuable
contribution to network science.

This paper addresses a specific network problem from the data fusion
perspective.  Over the past decade, there has been a great deal of
work on the {\em community detection} problem~\cite{For10}: discerning
how a graph's nodes are organized into ``communities.''  There is no
universally accepted definition of community structure: it can
correspond to some unobserved, ground-truth organizational structure;
it can refer to some attribute that nodes share that drives them to
``flock'' together~\cite{MSC01}; or communities can be defined as sets
of nodes more densely connected to each other than to the rest of the
graph.  Whatever the definition of community structure, it nearly
always results in communities being densely connected subsets of nodes
(the Newman--Leicht algorithm being a notable
exception~\cite{NeLe07}).  In practice, studies of community structure
in graphs (e.g.,~\cite{LKSF10}) define a community to be, in effect,
the output of a community detection algorithm.  Weighted and/or
directed edges are allowed in some methods, but accounting for more
general features on nodes and/or edges is problematic for network
research because this information tends to be domain-specific.

Community detection is nearly always formulated in terms an {\em
algorithm} which ingests a network and outputs some indication of its
community structure.  With a few exceptions, community detection
algorithms produce a single, hard-call result.  Most often this result
is a {\em partition} of nodes into non-overlapping communities, but a
few algorithms produce overlapping communities (e.g.,
CFinder~\cite{PDFV05}), and some produce a {\em dendrogram}---i.e., a
hierarchy of partitions~\cite{RoBe11}.  The dominant framework for
finding the best partition of nodes is to specify some {\em quality
function} of a partition relative to a graph and seek to maximize it.
Methods that maximize {\em modularity}~\cite{NeGi04} (explicitly or
implicitly) are among the most numerous and successful today.

From a data fusion perspective, however, it is important to assess the
uncertainty associated with community detection.  Quality functions
such as modularity are only motivated by intuition or physical
analogy, whereas probability is the language of logical reasoning
about uncertainty~\cite{Jay03}.  The reason principled fusion of
disparate data types is possible is that one can posit an underlying
model for reality, along with measurement models that specify the
statistics of how this reality is distorted in the data.  One can then
update one's prior distribution on reality to a posterior via Bayesian
inference~\cite{Var96}.

There are some methods that formulate community detection as an {\em
inference} problem: a prior distribution over all possible community
structures is specified, along with a likelihood function for
observing a graph given a community structure.  Hastings, for
example, formulated the community detection problem in terms of a {\em
Potts model} that defines the Hamiltonian $H$ for a given
graph--partition pair, and then converted this to a probability
proportional to $e^{-\beta H}$~\cite{Has06}.  Minimizing $H$ therefore
yields the MAP (Maximum {\it A posteriori} Probability) partition for given
structural parameters of the Potts model.  Hofman and Wiggins extended
this approach by integrating the structural parameters against a
prior~\cite{HoWi08}.  In both cases, if all one does with the
posterior probability distribution is locate its maximum, then it
becomes, in effect, just another quality function (albeit a principled
one).  On the other hand, the entire posterior distribution is vast,
so one cannot simply return the whole thing.  The question, then, is
what such a probability distribution is good for.

Clauset et al. made greater use of the posterior distribution by
devising a Monte Carlo method for generating dendrograms, and using it
to estimate the probabilities of missing links~\cite{CMN08}.
Reichardt and Bornholdt employed a similar Monte Carlo
method to estimate the pairwise co-membership probabilities $\pvw$ between
nodes~\cite{ReBo04,ReBo06}, where $\pvw$ is defined to be the
probability that nodes $v$ and $w$ are in the same community.
The set of all $\pvw$ is much smaller than the full posterior
distribution, and thus provides a useful, if incomplete, summary of the
uncertainty information.  It is expensive to compute exactly,
however.  Therefore we will derive an accurate approximation with
which to summarize uncertainty information for community
structure more efficiently than Monte Carlo methods.

A key benefit of retaining uncertainty information is that it enables
principled tracking~\cite{SBC99}.  We may track time-varying
communities in time-varying graph data by deriving an efficient
Bayesian filter for tracking time-varying communities from
time-varying graph data.  The term ``filter'' is somewhat strange in
this context: the original, signal-processing context of filters
(e.g., the Wiener filter~\cite{Wie49}) was that of algorithms which
filter out noise in order to highlight a desired signal.  The Kalman
filter changed this framework to one of distinct {\em state} and {\em
measurement} spaces~\cite{Kal60}.  This was soon generalized to the
concept of a {\em Bayesian filter}~\cite{Jaz70}.  To develop a
Bayesian filter, one constructs (a) an evolution model over a state
space that specifies the probability distribution of the state at some
future time given its value at the current time, and (b) a measurement
model that specifies the probability distribution of the current
measurement given the current state.  Thus, despite the connotations
of the word ``filter,'' a Bayesian filter can have quite different
state and measurements spaces.  To track communities, a model for the
co-evolution of graphs and community structure will be constructed,
and the measurement model will be that only the graph component of
the state is observable.

In Section~\ref{sec:staticExact} we derive exact inference equations
for the posterior probabilities of all possible community structures
for a given graph.  This result is essentially the same as can be
found elsewhere (e.g.,~\cite{Has06,HoWi08,KaNe11}), but is included
here in order to introduce notation and clarify subsequent material.
In Section~\ref{sec:staticApprox} we derive an approximation of the
co-membership probabilities $\pvw$ based on using only the most
important information from the graph.  The $\pvw$ matrix provides the
uncertainty information that the usual hard-call algorithms lack.  In
Section~\ref{sec:staticResults} we demonstrate that the $\pvw$
approximation is accurate and also surprisingly efficient: despite the
fact that it provides so much information, it is significantly faster
than the current, state-of-the-art community detection algorithm
(Infomap~\cite{RoBe08,LaFo09}).  We also demonstrate the uses for this
alternative or supplemental form of community detection, which are
embodied in the software IGNITE (Inter-Group Network Inference and
Tracking Engine).  One benefit of maintaining uncertainty information
is that it allows principled tracking.  In
Section~\ref{sec:dynamicExact} we present a continuous-time Markov
process model for the time-evolution of both the community structure
and the graph.  We then derive an exact Bayesian filter for this
model.  The state space for this model is far too large to use in
practice, so in Section~\ref{sec:dynamicApprox} we discuss efficient
approximations for the exact filter.  The community tracking material
is less developed than the corresponding detection material: there are
several issues that must be resolved to develop accurate, efficient
tracking algorithms.  However, we believe that the principled
uncertainty management of the data fusion approach provides a
framework for the development of more reliable, robust community
tracking methods.

\section{Community Detection:  Exact Equations}
\label{sec:staticExact}

Suppose that out of the space $\bK$ of all possible networks on $n$
nodes we are given some particular network $\kappa$.  If we have some
notion of a space $\bPhi$ of all possible ``community structures'' on
these $n$ nodes, then presumably the network $\kappa$ provides some
information about which structures are plausible.  One way to
formalize this notion is to stipulate a quality function $Q: \bK
\times \bPhi \ra \mathbb{R}$ that assigns a number to every
network--structure pair $(\kappa,\phi)$.  It would be natural, for
example, to define quality as a sum over all node pairs $\{v,w\}$ of
some metric for how well the network and community structure agree at
$\{v,w\}$.  That is, in the network $\kappa$, if $\{v,w\}$ is a link (or a
``strong'' link, or a particular kind of link, depending on
what we mean by ``network''), then it should be rewarded if $\phi$
places $v$ and $w$ in the same community (or ``nearby'' in community
space, or in communities consistent with the observed link
type, depending on what we mean by ``community structure'').
Modularity is a popular, successful example of a quality
function~\cite{NeGi04}.  Quality functions are easy to work with
and can be readily adapted to novel scenarios.  However, the price of
this flexibility is that unless one is guided by some additional
structure or principle, the choice of quality function is essentially
{\it ad hoc}.  In addition, the output of a quality function is a number
that provides nothing beyond an ordering of the community structures
in $\bPhi$.  The ``quality'' itself has little meaning.

One way to give quality functions additional meaning is to let them
represent an {\em energy}.  In this case, the quality function may be
interpreted as a Hamiltonian.  The qualities assigned to various
community structures are no longer arbitrary scores in this case:
meaningful {\em probabilities} can be assigned to community structures
can be computed from their energies.  The language of statistical
physics reflects the dominance of that field in network
science~\cite{For10}, but from a fusion standpoint it is more natural
to dispense with Hamiltonians and work directly with the
probabilities.  A probabilistic framework requires {\em models}: these
necessarily oversimplify real-world phenomena, and one could argue
that specifying a model is just as arbitrary as specifying a quality
function directly.  However, the space of probabilistic models is much
more constrained than the space of quality functions, and, more
importantly, formulating the problem in terms of a formal probability
structure allows for the meaningful management of uncertainty.  For
this reason, modularity and other quality functions tend to be re-cast
in terms of a probability model when possible.  For example, the
modularity function of Newman and Girven~\cite{NeGi04} was generalized
and re-cast as the Hamiltonian of a Potts model by Reichardt and
Bornholdt~\cite{ReBo06}, while Hastings demonstrated that this is
essentially equivalent to inference (i.e., the direct manipulation of
probability)~\cite{Has06}.

A probabilistic framework for this community structure problem
involves random variables $\Kappa$ for the graph and $\Phi$ for the
community structure.  We require models for the prior probabilities
$\pr(\Phi = \phi)$ for all $\phi \in \bPhi$ and for the conditional
probability $\pr(\Kappa = \kappa | \Phi = \phi)$ for all $\kappa \in
\bK$ and $\phi \in \bPhi$.  (We will typically use less formal
notation such as $\pr(\phi)$ and $\pr(\kappa|\phi)$ when convenient.)
Bayes' theorem then provides the probability $\pr(\Phi = \phi| \Kappa
= \kappa)$ of the community structure $\phi$ given the graph data
$\kappa$.  The models $\pr(\phi)$ and $\pr(\kappa|\phi)$ typically
have unknown input parameters $\params$, so that the probability
given by Bayes' theorem could be written $\pr(\phi|\kappa,\params)$.
This must be multiplied by some prior probability $\pr(\params)$
over the parameter space and integrated out to truly
give~$\pr(\phi|\kappa)$~\cite{HoWi08}.  A simpler, but non-rigorous,
alternative to integrating the input parameters against a prior is to
{\em estimate} them from the data.  This can be accurate when they are
strongly determined by the graph data: i.e., when
$\pr(\params|\kappa)$ is tightly peaked.  The issue of integrating
out input parameters will be addressed in
Section~\ref{sec:staticApprox}, but for now we will not include them
in the notation.

Section~\ref{sec:block} will derive $\pr(\kappa|\phi)$ using
a {\em stochastic blockmodel}~\cite{DBF05} with multiple link types for $\kappa$.  In
Section~\ref{sec:planted}, this will be simplified to the special case
of a {\em planted partition}~\cite{CoKa01} model in which links are
only ``on'' or ``off.''  

\subsection{Stochastic blockmodel case}
\label{sec:block}

Let $m$ denote the number of {\em communities}, and $r$ be the number
of {\em edge types}.  The notation $[p]$ will denote the set of
integers $\{1,2,\dots,p\}$, and $[p]_0$ denote the zero-indexed set
$\{0,1,\dots,p-1\}$.  We will let $[n]$ denote the set of nodes;
$[m]$, the set of communities; and $[r]_0$, the set of edge types.
Let $S^{\{2\}}$ denote the set of (unordered) pairs of a set $S$ so
that $\eset$ denotes the set of node pairs.  It is convenient to
consider $\eset$ to be the set of edges: because there are an
arbitrary number of edge types $r$, one of them (type $k = 0$)
can be considered ``white'' or ``off.''  Thus, all graphs have
$N \doteq n(n-1)/2$ edges, but in sparse graphs most of these are the trivial
type $k = 0$.

The community structure will be specified by a {\em community assignment}
$\phi: [n] \ra [m]$, i.e., a function that maps every node $v \in [n]$
to a community $\phi(v) \in [m]$.  The graph will be specified as a
function $\kappa: \eset \ra [r]_0$, which maps every edge $e \in \eset$
to its type $\kappa(e) \in [r]_0$.  (This unusual notation $\kappa$ will
be replaced with the more usual $G$ when dealing with the $r=2$ case:
i.e., when there is only edge type aside from ``off.'')

The {\em stochastic blockmodel} $H(n,\bp,\bQ)$ is parametrized by the
the number of nodes $n$, the stochastic $m$-vector $\bp$, and a
collection $\bQ$ of stochastic $r$-vectors $\bq_{ij}$~\cite{KaNe11}.  Here
``stochastic $m$-vector'' simply means a vector of length $m$ whose
components are non-negative and sum to one.  The vector $\bp$
comprises the prior probabilities $p_i$ of a node belonging to the
community $i \in [m]$---the communities for each node are drawn independently.
For $1 \le i \le j \le m$, the vector $\bq_{ij}$ comprises the
probabilities $q_{ij,k}$ of an edge between nodes in communities $i$ and
$j$ being of type $k$---the types of each edge are drawn independently
once the communities of the nodes are given.  (For $i > j$, let $\bq_{ij} =
\bq_{ji}$: i.e., the edges are undirected.)  The model $H(n,\bp,\bQ)$
defines the random variables $\Phi$ and $\Kappa$ whose instances are
denoted $\phi$ and $\kappa$, respectively.  The derivation of $\pr(\Phi =
\phi | \Kappa = \kappa)$ proceeds in six steps.

\uhead{Step 1.}  The probability that a node $v$ belongs to the
community $\phi(v)$ is, by definition,
\begin{equation}
\pr(\Phi(v) = \phi(v)) = p_{\phi(v)}.
\end{equation}

\uhead{Step 2.}  The probability that an instance of $\Phi$ is the community
assignment $\phi$ equals
\begin{equation}
\pr(\Phi = \phi) = \prod_{v=1}^n p_{\phi(v)}
\end{equation}
because the communities of each node are selected independently.

\uhead{Step 3.}  For a fixed value $\phi$ of $\Phi$, the probability that the
edge $e = \{v,w\}$ has type $\kappa(e)$ is, by definition,
\begin{equation}
\pr(\Kappa(e) = \kappa(e) | \Phi = \phi) = q_{\phi(v),\phi(w),\kappa(e)}.
\end{equation}

\uhead{Step 4.}  For a fixed value of $\phi$ of $\Phi$, the
probability that an instance of $\Kappa$ is the graph $\kappa$ equals
\begin{equation}
\pr(\Kappa = \kappa | \Phi = \phi) = \prod_{e \in \eset} q_{\phi(e_1),\phi(e_2),\kappa(e)},
\end{equation}
because the types of each edge are selected independently given
$\phi$.

\uhead{Step 5.}  The probability of a specific assignment $\phi$ and
graph $\kappa$ equals
\begin{equation}
\pr(\Phi = \phi, \Kappa = \kappa) = \prod_{v=1}^n p_{\phi(v)} \prod_{e
  \in \eset} q_{\phi(e_1),\phi(e_2),\kappa(e)}, \label{eq:prphikap}
\end{equation}
because $\pr(\phi,\kappa) = \pr(\kappa | \phi) \pr(\phi)$.

\uhead{Step 6.} Finally, the posterior probability of $\Phi = \phi$ for a given
graph $\kappa$ is
\begin{equation}
\pr(\Phi = \phi | \Kappa = \kappa) \propto \pr(\Phi = \phi, \Kappa =
\kappa), \label{eq:prphiGkap}
\end{equation}
where the constant of proportionality is $1/\pr(\Kappa = \kappa)$.

\subsection{Planted partition case}
\label{sec:planted}

In many applications one does not have any {\it a priori} knowledge
about specific communities.  In such cases, the community labels $[m]$ are
arbitrary: the problem would be unchanged if the communities were labeled
according to another permutation of $[m]$.  Thus, if one has a prior
distribution over $\bp$ and $\bQ$ (as in~\cite{HoWi08}), then that
distribution must be invariant under permutations of $[m]$.  In the
case of fixed input parameters $\bp$ and $\bQ$, this translates to
$\bp$ and $\bQ$ themselves being invariant under permutations.  Making
this simplification, and considering only $r = 2$ edge types (``off''
($k=0$) and ``on'' ($k=1$)) yields the special case called the {\em
planted partition} model~\cite{CoKa01}.  In this case, symmetry
implies that $p_i = 1/m$ for all $i \in [m]$, and that $q_{ij,1} =
p_I$ for $i = j$ and $q_{ij,1} = p_O$ for $i \ne j$.  Here $p_I$
denotes the edge probability between nodes in the same community, and
$p_O$, the edge probability between nodes in different communities.  Thus,
the $m + m(m+1)(r-1)/2$ input parameters of $H(n,\bp,\bQ)$ reduce to
only four to give the planted partition model $H(n,m,p_I,p_O)$.

Having only two edge types suggests using the standard notation $G$ to
denote a graph, with $E(G)$ denoting the set of (``on'') edges.  The
symmetry of the community labels implies that $\pr(\phi | \kappa)$ is
invariant under permutations of $[m]$, so that is more efficient to
formulate the problem in terms of a {\em partition} $\pi$ of the nodes
into communities rather than $\phi$ (because partitions are orbits of community
assignments under permutations of $[m]$).  We may then
replace~\eqref{eq:prphikap} by
\begin{equation}
\pr(\pi, G) = \frac{(m)_{|\pi|}}{m^n} p_I^{e_I} (1-p_I)^{\tle_I}
p_O^{e_O} (1-p_O)^{\tle_O}. \label{eq:prpiG}
\end{equation}
Here $|\pi|$ denotes the number of (non-empty) communities in the partition
$\pi$, and $(m)_k$ denotes the falling factorial $m!/(m-k)!$, which
counts the number of assignments $\phi$ represented by the equivalence
class $\pi$.  The number of edges between nodes in the same community is
denoted $e_I(G)$ (abbreviated to $e_I$ in~\eqref{eq:prpiG}), and the
number of non-edges (or ``off'' edges) between nodes in the same community
is denoted $\tle_I(G)$.  The analogous quantities for nodes in
different communities are $e_O(G)$ and $\tle_O(G)$.  The posterior
probability $\pr(\pi | G)$ is proportional to $\pr(\pi,G)$.

\section{Community Detection:  Approximate Methods}
\label{sec:staticApprox}

Community detection methods that employ quality functions return hard
calls:  an optimization routine is applied to determine the community
structure that maximizes the quality $f(\kappa,\phi)$ over all $\phi
\in \bPhi$ for a given graph $\kappa$.  There is little else one can
do with a quality function:  one can return an ordered list of the
$k$-best results, but a probability framework is required to interpret
the relative likelihoods of these.

In contrast, the formulas~\eqref{eq:prphikap} and~\eqref{eq:prpiG}
provide the information necessary to answer any statistical question
about the community structure implied by $\kappa$.  Unfortunately, an
algorithm that simply returns the full distribution is grossly
impractical.  The number of partitions of $n$ nodes is the {\em Bell
number} $B(n)$, which grows exponentially with $n$: e.g., $B(60)
\approx 10^{60}$.  What, then, are these probabilities good for?  One
answer is that the formula for posterior probability can be used as a
(more principled) quality function~\cite{Has06}.  Another is that
Monte Carlo methods can be used to produce a random sample of
solutions~\cite{ReBo04,CMN08}.  These random samples can be used to
approximate statistics of $\Phi$.  In this section we will consider
how such statistics might be computed directly.

\subsection{Stochastic blockmodel}
\label{sec:staticApproxBlock}

The most natural statistical question to ask is this:  what is the
probability that a node $v$ is in community $i$?  We may express this
probability as $p_i^v \doteq \pr(\Phi(v) = i | \Kappa = \kappa)$, where the
dependence on the graph $\kappa$ is suppressed from the notation.
For the model $H(n,\bp,\bQ)$, we may compute $p_i^v$ from~\eqref{eq:prphikap}:
\begin{equation}
p_i^v = \frac{1}{\pr(\kappa)} \sum_{\phi \in \bPhi \atop \phi(v)=i} \prod_{v=1}^n p_{\phi(v)} \prod_{e
  \in \eset} q_{\phi(e_1),\phi(e_2),\kappa(e)}. \label{eq:piv}
\end{equation}
Unfortunately, this exact expression does not appear to simplify in
any significant way.  (Ironically, its dynamic counterpart {\em does}
simplify:  cf. Section~\ref{sec:dynamicApprox}.)

A strategy for approximating $p_i^v$ is to use only the most relevant
information in the graph.  For example, we could divide the edges
into two classes:  those that contain $v$ and those that do not.
Edges in the former class have more direct relevance to the question
of which community $v$ belongs to.  If we let $\kappa_v$ denote the
restriction of the graph $\kappa$ to edges containing $v$, and
$\Kappa_v$ be the corresponding random variable, then we may
approximate $p_i^v$ by $\tilde{p}_i^v \doteq \pr(\Phi(v) = i | \Kappa_v =
\kappa_v)$.  By Bayesian inversion this equals
\begin{equation}
\begin{split}
\tilde{p}_i^v &\propto \pr(\Phi(v) = i) \pr(\Kappa_v = \kappa_v | \Phi(v) = i) \\ 
&= p_i \prod_{x \neq v} \pr(\Kappa(\{v,x\}) = \kappa(\{v,x\}) |
\Phi(v) = i) = p_i \prod_{x \neq v} \sum_{j=1}^m p_j
q_{ij,\kappa(\{v,x\})}.
\end{split} \label{eq:ptiv}
\end{equation}
This equation exploits the statistical distribution of edge types
that tend to emanate from a given community:  if $\bQ$ is such that
this information is distinctive, then~\eqref{eq:ptiv} will perform
well.  However, because it assesses each node in isolation, it does
not exploit network structure and will not perform well when $\bQ$
fails to produce distinctive edge-type distributions.

If there were multiple, conditionally independent graph snapshots
for a given ground-truth $\phi$, then one could replace $p_i$ with
$\tilde{p}_i^{v-}$ in~\eqref{eq:ptiv}, and $p_j$ with
$\tilde{p}_j^{x-}$, to get an updated value $\tilde{p}_i^{v+}$.  One
could initialize these values $\tilde{p}_i^{v-}$ to the prior $p_i$
and apply the update equation for each graph snapshot $\kappa$: this
would introduce communication between the results for individual nodes
and thus exploit network structure.  The approach in
Section~\ref{sec:dynamicApprox} is a more sophisticated version of
this, which allows the temporal sequence of graphs to be correlated
and nodes to move between communities.

To derive useful probabilistic information that exploits network structure
rather than just the statistical characteristics of edge-type
distributions we turn to the second-order statistics $p_{ij}^{vw} \doteq
\pr(\Phi(v) = i, \Phi(w) = j | \Kappa = \kappa)$.  To approximate
this, we may divide the edges into three classes:  the edge $\{v,w\}$,
the edges containing either $v$ or $w$ (but not both), and the edges
containing neither.  One gets a rather trivial approximation using
only the single edge $\{v,w\}$, but using edges from the first two
classes yields the approximation $\tilde{p}_{ij}^{vw} \doteq \pr(\Phi(v) =
i, \Phi(w) = j | \Kappa_v = \kappa_v, \Kappa_w = \kappa_w)$.  This
quantity has a formula similar to~\eqref{eq:ptiv}:
\begin{equation}
\begin{split}
\tilde{p}_{ij}^{vw} &\propto \pr(\Phi(v) = i, \Phi(w) = j)
\pr(\Kappa_v = \kappa_v, \Kappa_w = \kappa_w | \Phi(v) = i, \Phi(w) = j) \\ 
&= p_i p_j \pr(\Kappa(\{v,w\}) = \kappa(\{v,w\}) | \Phi(v) = i, \Phi(w) = j) \times \\
& \qquad \prod_{x \neq v,w} \pr(\Kappa(\{v,x\}) = \kappa(\{v,x\}), \Kappa(\{w,x\}) = \kappa(\{w,x\}) |
\Phi(v) = i, \Phi(w) = j) \\
&= p_i p_j q_{ij,\kappa(\{v,w\})} \prod_{x \neq v,w} \sum_{k=1}^m p_k q_{ik,\kappa(\{v,x\})} q_{jk,\kappa(\{w,x\})}.
\end{split} \label{eq:ptijvw}
\end{equation}
(The version that uses only the single edge $\{v,w\}$ as evidence is
given by omitting the final product in~\eqref{eq:ptijvw}.)  This
formula provides important statistical information even when $\bQ$ is
completely symmetric.  Indeed, to exploit $\tilde{p}_{ij}^{vw}$ it is
simpler to work with the symmetric case.

\subsection{Planted partition model}
\label{sec:staticApproxPlanted}

When $\bp$ and $\bQ$ are symmetric under permutations of $[m]$,
then~\eqref{eq:piv} reduces to $p_i^v = 1/m$ (and~\eqref{eq:ptiv} to
$\tilde{p}_i^v = 1/m$).  This is because in the symmetric case
$H(n,m,p_I,p_O)$ community labels have no meaning, so first-order
statistics become trivial.  The simplest, non-trivial quantities to
compute are the second-order statistics $p_{ij}^{vw}$.  In the
symmetric case, they reduce to the single probability $\pvw$ that $v$
and $w$ are in the same community: i.e., $\pvw \doteq \pr(\Phi(v) = \Phi(w)
| \Kappa = \kappa)$.  To compute $\pvw$ exactly requires a summation
over all partitions.  Reichardt and Bornholdt estimated the $\pvw$
matrix by a Monte Carlo sampling of the partition space, but this is
slow~\cite{ReBo06}.  Instead of this, we may approximate $\pvw$
directly by simplifying~\eqref{eq:ptijvw}.  This leads to fairly
simple expressions.  The meaning of these expressions is opaque,
however, when derived through straightforward mathematical
manipulations, which creates problems when trying to adapt the results
to engineering contexts.  Therefore we proceed along more general
lines to demonstrate which aspects of the partition--graph model lead
to which aspects of the resulting expressions.

Suppose instances of some random process $\mathcal{P}$ are
partition--graph pairs $(\pi,G)$ on $n$ nodes.  This process is not
necessarily $H(n,m,p_I,p_O)$: we will later take
$\mathcal{P}$ to be a somewhat more complex process in which the
parameters $m$, $p_I$, and $p_O$ are first drawn from some
distribution, and then an instance of $H(n,m,p_I,p_O)$ is generated.  Let
$M_{vw}$ be the indicator random variable for the event that $v$ and
$w$ are in the same community (i.e., $M_{vw} = 1$ when $v$ and $w$ are
in the same community, and 0 otherwise), and $\Kappa_{vw}$ be the
indicator random variable for the existence of an edge between $v$ and
$w$.  Now let $\kappa_{vw}$ indicate the presence or absence of the
edge $\{v,w\}$ in some given graph $G$ (i.e., $\kappa_{vw} = 1$ if
$\{v,w\}$ is an edge of $G$, and 0 otherwise).  Thus, the $\kappa_{vw}$
are data, rather than instances of $\Kappa_{vw}$.  We define $J_{vw}$
to be the indicator random variable for $\Kappa_{vw}$ agreeing with this
datum $\kappa_{vw}$ (i.e., $J_{vw} = 1$ if $\Kappa_{vw} = \kappa_{vw}$,
and 0 otherwise.  We may express $J_{vw}$ as
\begin{equation}
  J_{vw} \doteq 1 - \kappa_{vw} - (-1)^{\kappa_{vw}} \Kappa_{vw}. \label{eq:jdef}
\end{equation}
Now let $\Xi_{vw}$ be the indicator random variable for $\mathcal{P}$
agreeing exactly with $G$ on all edges containing $v$ and/or $w$.  We
may express this as
\begin{equation}
  \Xi_{vw} \doteq J_{vw} \prod_{x \neq v,w} J_{vx} J_{wx}. \label{eq:Xi}
\end{equation}

The approximation to $\pvw$ based on using only local graph
information may then be written
$\ptvw \doteq \pr(M_{vw} = 1 | \Xi_{vw} = 1) = \expt[M_{vw} | \Xi_{vw}
= 1]$.  This can be expressed as
\begin{equation}
\ptvw = \frac{\pr(\Xi_{vw} = 1 | M_{vw} = 1) \pr(M_{vw} = 1)}{\pr(\Xi_{vw} = 1)} =
\frac{\Lambda_{vw}}{\Lambda_{vw} + \expt[M_{vw}]^{-1} - 1}, \label{eq:ptvw}
\end{equation}
where the likelihood ratio $\Lambda_{vw}$ is given by
\begin{equation}
\Lambda_{vw} \doteq \frac{\expt[\Xi_{vw} | M_{vw} = 1]}{\expt[\Xi_{vw} | M_{vw} = 0]}. \label{eq:lambdavw}
\end{equation}

To evaluate $\expt[\Xi_{vw} | M_{vw}]$ we would like to
use~\eqref{eq:Xi}, requiring that $\mathcal{P}$ have suitably
favorable properties.  If $\mathcal{P} = H(n,m,p_I,p_O)$, then the
random variables $J_{vw}$, and each of the $J_{vx} J_{wx}$ for $x \neq
v,w$ are conditionally independent given $M_{vw}$.  E.g., if $M_{vw} =
1$ (i.e., $v$ and $w$ are in the same community), then $K_{vw} = 1$
with probability $p_I$, independent of the values of any other
$K_{xy}$.  However, if $\mathcal{P}$ is a process in which a parameter
vector $\params$ is first drawn from some distribution, and then a
draw is made from some process $\mathcal{P(\params)}$, then
assumption of conditional independence is far too restrictive.  In
such a case the existence of many edges elsewhere in the graph would
suggest a large value of a parameter like $p_I$, and hence a larger
value of $K_{vw}$, so this random variable would not be conditionally
independent of the other $K_{xy}$ given $M_{vw}$.

This problem is easily overcome, however.  We simply decompose the
expected value into the conditional expectation for a specified value of
$\params$, followed by an expectation over $\params$.  E.g., we write
$\expt[\Xi_{vw} | M_{vw}]$ as
\begin{equation}
  \expt[\Xi_{vw} | M_{vw}] = \expt_{\params}[\expt[\Xi_{vw} | M_{vw},\params]].
\label{eq:eXi}
\end{equation}
We then stipulate that $J_{vw}$ and each of the
$J_{vx} J_{wx}$ for $x \neq v,w$ are conditionally independent given
$M_{vw}$ and $\params$.  Then
\begin{equation}
  \expt[\Xi_{vw} | M_{vw},\params] = \expt[J_{vw} | M_{vw},\params] \prod_{x
    \neq v,w} \expt[J_{vx} J_{wx} | M_{vw},\params]. \label{eq:XiDecomp}
\end{equation}
We may express the factors in the product in terms of a covariance:
\begin{equation}
\expt[J_{vx} J_{wx} | M_{vw},\params] =
  \expt[J_{vx} | M_{vw},\params] \expt[J_{wx} | M_{vw},\params] +
  \mathrm{Cov}(J_{vx},J_{wx} | M_{vw},\params).
\end{equation}
We make the further assumption that $K_{vx}$ is conditionally
independent of $M_{vw}$ given $\params$ (which, again, holds for
$H(n,m,p_I,p_O)$).  Then, using~\eqref{eq:jdef} we have
\begin{equation}
\expt[J_{vx} J_{wx} | M_{vw},\params] =
  \expt[J_{vx} | \params] \expt[J_{wx} | \params] +
  (-1)^{\kappa_{vx} + \kappa_{wx}} \mathrm{Cov}(K_{vx},K_{wx} | M_{vw},\params).
\end{equation}

We introduce the following notation
\begin{alignat}{2}
\mu &\doteq \expt[M_{vw} | \params] , & \qquad \psi^+ &\doteq
\mathrm{Cov}(K_{vx},K_{wx} | M_{vw} = 1,\params), \label{eq:mudef} \\
\delta &\doteq \expt[K_{vx} | \params], &
\psi^- &\doteq \mathrm{Cov}(K_{vx},K_{wx} | M_{vw} = 0,\params).
\end{alignat}
In this symmetric scenario all quantities are invariant under node
permutations.  Thus $\mu$ is the probability that two randomly chosen nodes
are in the same community (for fixed parameters $\params$), and $\delta$
is the probability that two random chosen nodes have an edge between
them.  We write $\expt[J_{vx} J_{wx} | M_{vw},\params]$ in terms of
these quantities:
\begin{equation}
\expt[J_{vx} J_{wx} | M_{vw},\params] =
\begin{cases}
(1-\delta)^2 + \psi & \text{if} \; \kappa_{vx}+\kappa_{wx} = 0, \\
\delta (1-\delta) - \psi & \text{if} \; \kappa_{vx}+\kappa_{wx} = 1, \\
\delta^2 + \psi & \text{if} \; \kappa_{vx}+\kappa_{wx} = 2,
\end{cases}
\end{equation}
where
\begin{equation}
\psi \doteq
\begin{cases}
\psi^- & \text{if} \; M_{vw} = 0, \\
\psi^+ & \text{if} \; M_{vw} = 1.
\end{cases}
\end{equation}
We may use this to express~\eqref{eq:XiDecomp} as
\begin{equation}
  \expt[\Xi_{vw} | M_{vw},\params] = \expt[J_{vw} |
  M_{vw},\params] f(\delta,\psi), \label{eq:Xif}
\end{equation}
where
\begin{equation}
f(\delta,\psi) \doteq
\big((1-\delta)^2 + \psi\big)^{n_0}
\big(\delta (1-\delta) - \psi\big)^{n_1}
\big(\delta^2 + \psi\big)^{n_2}.
\end{equation}
Here $n_j$ denotes the number of nodes (aside from $v$ and $w$)
adjacent to exactly $j$ of $\{v,w\}$, and $n_0 + n_1 + n_2 = n-2$.

Now to compute $\expt[\Xi_{vw} | M_{vw}]$ we substitute~\eqref{eq:Xif}
into~\eqref{eq:eXi}.  To evaluate the expectation
$\expt_{\params}[\cdot]$ of~\eqref{eq:Xif} requires a specific random
graph model $\mathcal{P}$.  We will use the following $\mathcal{P}$:
we will select the number of communities $m$ in a manner to be
discussed below, and select $p_I$ and $p_O$ uniformly from $0 \le p_O
\le p_I \le 1$.  Then we shall make a draw from $H(n,m,p_I,p_O)$ to
generate a partition--graph pair $(\pi,G)$.  For this model we have
\begin{equation}
\mu = 1/m, \quad \text{and} \quad \delta = \mu p_I + (1-\mu) p_O,
\end{equation}
as well as
\begin{alignat}{3}
\psi^+ &= (\delta - p_O)(p_I - \delta) &&= \mu(1-\mu)(p_I-p_O)^2 &&
\ge 0, \;\; \text{and} \label{eq:psip}\\
\psi^- &= -(\delta - p_O)^2 &&= -\mu^2(p_I-p_O)^2 && \le 0. \label{eq:psim}
\end{alignat}
Finally, the leading factor in~\eqref{eq:Xif} is
\begin{alignat}{4}
 \expt[J_{vw}| M_{vw} = 1,\params] &= p_I& \;\text{if} \;\;\kappa_{vw} &= 1,&\quad 
 \expt[J_{vw}| M_{vw} = 1,\params] &= 1-p_I& \;\text{if} \;\;\kappa_{vw} &= 0, \label{eq:eJ1} \\ 
 \expt[J_{vw}| M_{vw} = 0,\params] &= p_O& \;\text{if} \;\;\kappa_{vw} &= 1,&\quad 
 \expt[J_{vw}| M_{vw} = 0,\params] &= 1-p_O& \;\text{if} \;\;\kappa_{vw} &= 0 \label{eq:eJ0} .
\end{alignat}

We may split the expectation $\expt_{\params}[\cdot]$ into an integral
over $p_I$ and $p_O$ followed by an expectation with respect to $m$.
Then~\eqref{eq:eXi} becomes
\begin{equation}
  \expt[\Xi_{vw} | M_{vw}] = \expt_{m}\left[
2 \int_0^1 \int_0^{p_I} \expt[J_{vw} |
  M_{vw},\params] f(\delta,\psi) \, dp_O dp_I\right]. \label{eq:eXpIpO}
\end{equation}
We may change coordinates from $(p_O,p_I)$ to $(\delta,\psi^+)$ for $M_{vw} =
1$ and to $(\delta,\psi^-)$ for $M_{vw} = 0$.  This introduces
complications due to Jacobians and complicated regions of integration
$R^+$ and $R^-$, but it is helpful to be in the natural coordinate
system of $f$:
\begin{align}
\expt[\Xi_{vw} | M_{vw} = 1] &= \expt_{m}\left[\iint_{R^+} \frac{\expt[J_{vw} |
  M_{vw}=1,\params]}{\sqrt{\mu (1-\mu) \psi^+}} f(\delta,\psi^+) \, d\delta d\psi^+ \right], \label{eq:eX1}\\
\expt[\Xi_{vw} | M_{vw} = 0] &= \expt_{m}\left[\iint_{R^-} \frac{\expt[J_{vw} |
  M_{vw}=0,\params]}{\mu\sqrt{-\psi^-}} f(\delta,\psi^-) \, d\delta d\psi^- \right] \label{eq:eX0}.
\end{align}
In the $M_{vw} = 1$ case, the range $0 \le p_O \le p_I \le 1$ is transformed into the following region $R^+$:
$\psi^+ = 0 \;\text{to}\; \delta^2 (1-\mu)/\mu$ for $\delta = 0 \;\text{to}\; \mu$ and
$\psi^+ = 0 \;\text{to}\; (1-\delta)^2 \mu/(1-\mu)$ for $\delta = \mu \;\text{to}\; 1$.
Similarly, in the $M_{vw} = 0$ case it is transformed into the following region $R^-$:
$\psi^- = -\delta^2 \;\text{to}\; 0$ for $\delta = 0 \;\text{to}\; \mu$ and
$\psi^- = -(1-\delta)^2 (\mu/(1-\mu))^2 \;\text{to}\; 0$ for $\delta = \mu \;\text{to}\; 1$.
To compute~\eqref{eq:eX1} and~\eqref{eq:eX0} numerically one would transform
the expressions~\eqref{eq:eJ1} and~\eqref{eq:eJ0} into $(\delta,\psi)$ space,
although it seems to be more numerically stable to use the
expressions~\eqref{eq:psip} and~\eqref{eq:psim} in~\eqref{eq:eXpIpO}.
For small $n$, this numerical integration is feasible.  The following
example employs numerical integration for a dataset with $n = 34$ nodes.

\definecolor{pvwred}{rgb}{0.8,0.2,0.25}
\definecolor{pvwblue}{rgb}{0.25,0.2,0.8}

\begin{figure}
  \centering
    \centerline{\includegraphics[width=0.5\columnwidth]{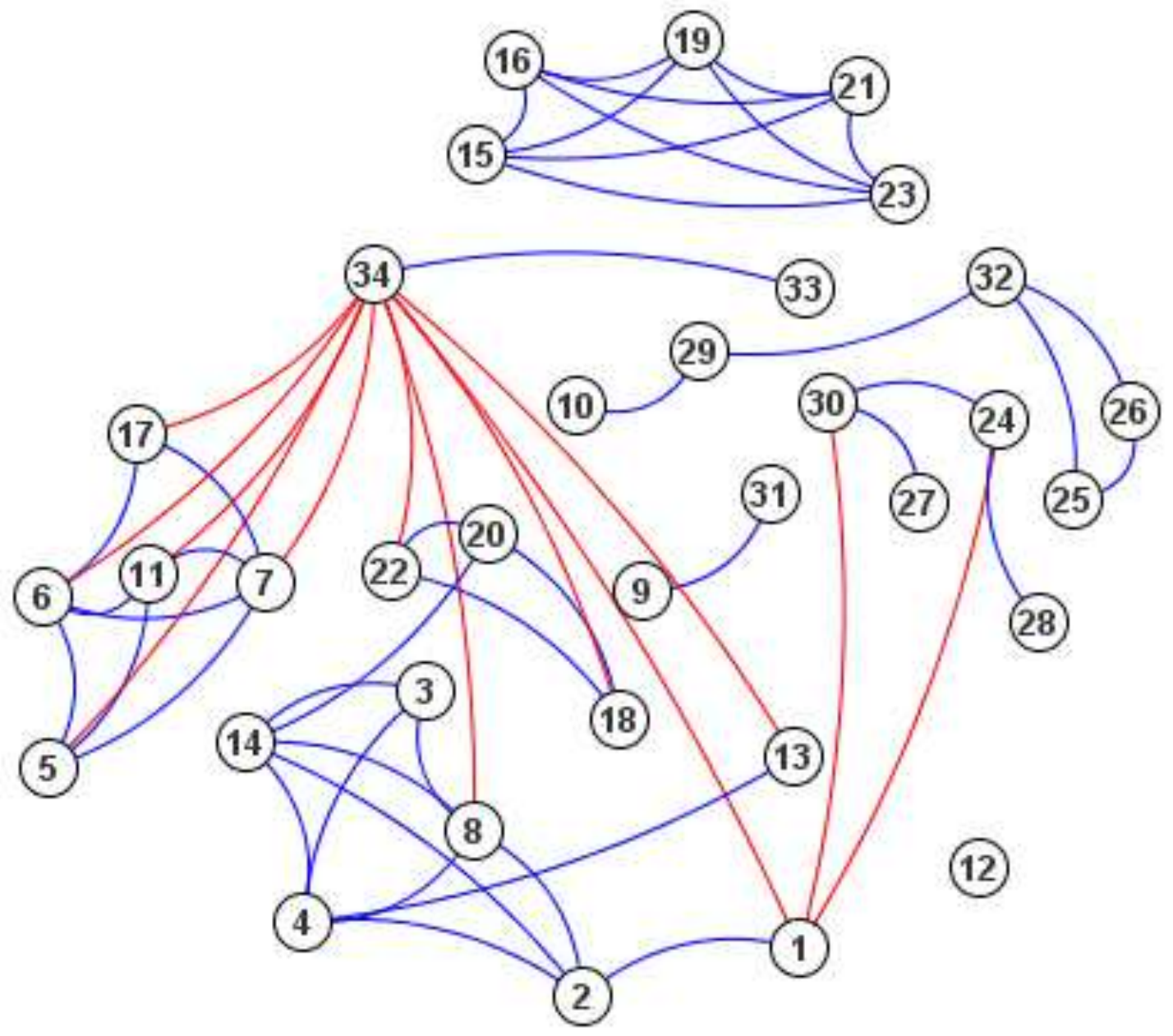}}

\caption{$\ptvw$ plot for Zachary's karate club:
{\color{pvwblue} blue}: $\ptvw \ge 60\%$;
{\color{pvwred} red}:  $\ptvw \le 1.8\%$
}
\label{fig:pvwkarate}
\end{figure}
Figure~\ref{fig:pvwkarate} shows which pairs of nodes are particularly
likely or unlikely to be in the same community for Zachary's
karate club data~\cite{Zac77}.  This is a social network of
34 members of a karate club at a university which split into two
communities. Nodes 4 and 8 are most likely, with $\tilde{p}^{\{4,8\}} = 98.8\%$, and nodes
1 and 34 least likely to be in the same community with $\tilde{p}^{\{1,34\}} =
0.65\%$.  Being adjacent does not guarantee a high value of $\ptvw$:
the node pairs $\{1,32\}$ and $\{14,34\}$ each have $\ptvw = 8.9\%$.
Nor is it necessary for nodes to be adjacent to have a high value of
$\ptvw$: among nodes 15, 16, 19, 21, and 23 $\ptvw = 84.5\%$ for every
pair, and $\tilde{p}^{\{8,14\}} = 96.1\%$.  Note that the node pairs
$\{8,14\}$ and $\{1,34\}$ are each non-adjacent, and each has four
common neighbors (i.e., $n_2 = 4$), but their values of $\ptvw$ differ
by a factor of 150 because of the degrees of the nodes involved.
Finally, although nodes 9 and 31 are in different ground-truth communities,
$\tilde{p}^{\{9,31\}} = 92.1\%$.

When numerical integration is not feasible, it is difficult to obtain
good asymptotic estimates as $n \ra \infty$, so we will resort to
heuristics.  The function $f$ has a global maximum which is
increasingly sharp as $n \ra \infty$.  This occurs at
\begin{equation}
  \label{eq:peak}
  \delta_p \doteq \frac{n_1 + 2n_2}{2(n-2)}, \quad
  \psi_p \doteq \frac{4 n_0 n_2 - n_1^2}{4 (n-2)^2}.
\end{equation}
If $\psi_p \ge 0$, then this peak lies within $R^+$ when $\psi_p /
(\psi_p + (1-\delta_p)^2) \le \mu \le \delta_p^2 / (\psi_p +
\delta_p^2).$ Assuming the expectation $\expt_m[\cdot]$ has
significant weight in this range, one can replace $f$ with a delta
function at $(\delta_p,\psi_p)$ to estimate $\expt[\Xi_{vw} | M_{vw} =
1]$ (i.e., parameter estimation is an appropriate approximation to the
full, Bayesian integration).  To estimate $\expt[\Xi_{vw} | M_{vw} =
0]$ in this case, one could make the same argument, but with the
maximum of $f(\delta,\psi^-)$ constrained to $\psi^- \le 0$.
Conveniently, this constrained maximum occurs at $(\delta_p,0)$.
The analogous argument works when $\psi_p \le 0$.  Therefore, if the
integrals in~\eqref{eq:eX1} and~\eqref{eq:eX0} contained nothing but
$f(\delta,\psi)$ we could approximate them by $f(\delta_p,\psi_p)$ or
$f(\delta_p,0)$ as appropriate.  Ignoring the expectation over $m$ as
well, we could substitute these expressions into~\eqref{eq:lambdavw}
to obtain the following approximation
\begin{equation}
  \label{eq:lambdatilde}
  \widetilde{\Lambda}_{vw} \approx \begin{cases}
    f(\delta_p,\psi_p)/f(\delta_p,0) & \mathrm{if} \; \psi_p \ge 0, \; \text{and}\\
    f(\delta_p,0)/f(\delta_p,\psi_p) & \mathrm{if} \; \psi_p < 0.
  \end{cases}
\end{equation}

We may decompose $\Lambda$ as $\Lambda = \widetilde{\Lambda} \Lambda_k$, where
$\Lambda_1$ encompasses all corrections to the crude approximation
$\widetilde{\Lambda}$ when there is an edge between $v$ and $w$, and
$\Lambda_0$ when there is no edge.  For this, we need to specify the
prior on $m$.  Here, we let $\log m$ vary uniformly from $\log 2$ to
$\log n$.  It is convenient to treat $m$ (or, equivalently, $\mu$) as a
continuum variable here to avoid the accidents of discreteness.  With
this prior, we may compute the value of $\expt[M_{vw}]$ required
in~\eqref{eq:ptvw} as
\begin{equation}
  \label{eq:mubar}
  \bar{\mu} = \expt[M_{vw}] = \frac{1/2 - 1/n}{\log(n/2)}.
\end{equation}
When $\psi_p$ differs greatly from 0, the $\widetilde{\Lambda}$ factor
is very large or small and dominates the correction term $\Lambda_k$.  Therefore
we seek to approximate the correction factor only in the critical case
$\psi_p = 0$.  Typically, real-world graphs are sparse, in which case
the lack of an edge between $v$ and $w$ decreases their co-membership
probability only slightly, but the presence of an edge greatly
enhances it.  Numerical experimentation confirms this intuition: the
correction factor $\Lambda_0$ due to the absence of an edge is roughly
constant, but the factor $\Lambda_1$ due to the presence of the edge
$\{v,w\}$ increases rapidly as $\delta_p$ decreases until it hits a
constant plateau (which varies with $n$):
\begin{align}
  \Lambda_0 &= \min(0.7197, 0.46 \, \delta_p^{-0.15}), \; \text{and} \label{eq:lam0} \\
  \Lambda_1 &= \min(0.5605\, n + 1.598, \delta_p^{-0.7}). \label{eq:lam1}
\end{align}
The four-digit coefficients in these formulas are obtained from an
asymptotic analysis of exact results obtained in the $n_1 = n_2 = 0$
case.  These exact results involve combinations of generalized
hypergeometric functions (i.e., ${}_p F_q(a;b;z)$), and are not
particularly enlightening, although they can be used to obtain
accurate coefficients, such as 0.56051044368284805729 rather than 0.5605
in~\eqref{eq:lam1}.

Putting the above together into~\eqref{eq:ptvw}, we obtain the following approximation to $\pvw$:
\begin{equation}
  \label{eq:phvw}
    \phvw =
      \frac{\Lambda_{\kappa_{vw}} \widetilde{\Lambda}}{\Lambda_{\kappa_{vw}} \widetilde{\Lambda} + \bar{\mu}^{-1} - 1}.
\end{equation}
This formula could certainly be improved.  It often yields results
such as $\phvw = 1 - 10^{-20}$: this figure might be accurate
given the model assumptions, but such certainty could never be
attained in the real world.  To make it more accurate a more
sophisticated model could be used, or the priors on $p_I$, $p_O$, and
$m$ could be matched more closely to reality.  Only limited
improvement is possible, however, because in reality multiple,
overlapping, fuzzily-defined community structures typically exist at
various scales, and it is unclear what $\pvw$ means in such a
context.  Certainly the integral approximations could be performed
more rigorously and accurately.  The broad outlines of the behavior
of $\pvw$ are captured in $\widetilde{\Lambda}$, $\Lambda_0$,
and $\Lambda_1$, however.  Finally, using only local evidence
constitutes a rather radical pruning of the information
in $G$.  However, it is because of this pruning that the
approximation~\eqref{eq:phvw} can be implemented so efficiently.

\section{Community Detection:  Results}
\label{sec:staticResults}

\begin{figure}
  \centering
    \centerline{\includegraphics[width=0.4\columnwidth]{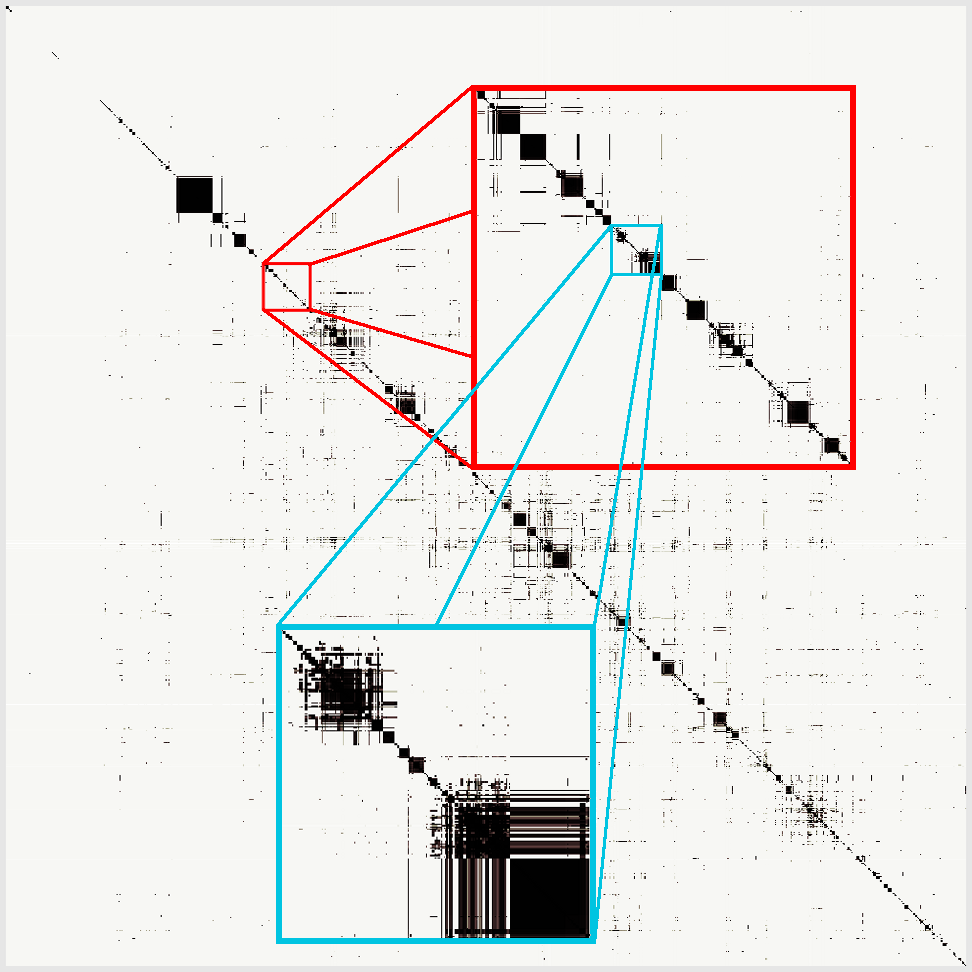}}
%
\caption{Visualization of $\phvw$ matrix for the Enron email network}
\label{fig:enron}
\end{figure}

Direct visualizations like Figure~\ref{fig:pvwkarate} are impractical
for larger networks.  It can be useful to use the blue edges in
Figure~\ref{fig:pvwkarate} (those with $\pvw$ above a certain
threshold) in place of a graph's edges in network algorithms (such as
graph layout): this is discussed in Section~\ref{sec:uses}.  However,
a more direct use of the co-membership matrix $\pvw$ for network
visualization is simply to plot the matrix itself with the values of
$\pvw \in [0,1]$ as intensities~\cite{ReBo04,ReBo06}.  An example of
this is shown in Figure~\ref{fig:enron}, using the
approximation~\eqref{eq:phvw}, for the Enron email communication
network, which has 36,692 nodes and 367,662 edges~\cite{LKF05}.  The
insets depict the hierarchical organization of community structure in
networks~\cite{CMN08,LFK09}:  communities with various structures exist at
all scales.  Although the model $H(n,m,p_I,p_O)$ does not account for
hierarchical structure, a benefit of integrating over the number of
communities $m$ (rather than estimating it) is that this accounts for
co-membership at different scales.

\subsection{Accuracy}
\label{sec:accuracy}

One may rightly question whether the approximation $\phvw$ is
accurate, given the modeling assumptions and approximations that it is
based on.  To address this, we observe that the values of $\pvw$ may
be used to define a certain family expected utility functions
(parameterized by a threshold probability $\theta$:
cf.~\eqref{eq:fullutilresult} in Appendix~\ref{sec:utility}), and
optimizing this expected utility yields a traditional community
detection algorithm.  Because a great many community detection
algorithms have been developed, one can assess the quality of the
approximation $\phvw \approx \pvw$ by comparing the performance of the
resulting community detection algorithm to those in the literature.

The most comprehensive comparison to date is based on the LFR
benchmark graphs which have power-law distributions both on degree and
on community size~\cite{LaFo09}.  The conclusion is that all
algorithms prior to 2008 were eclipsed by a set of three more recent
algorithms: the Louvain algorithm~\cite{BGLL08}, the
Ronhovde--Nussinov (RN) algorithm~\cite{RoNu09}, and
Infomap~\cite{RoBe08}.  Infomap performed somewhat better than RN, and
both somewhat better than Louvain, but all three were much better than
the previous generation.  Figure~\ref{fig:GFcompare} compares our
algorithm to the Infomap, RN, and Louvain algorithms, and to the other
algorithms tested in~\cite{LaFo09}.  (This figure is a correction of Figure 3
from~\cite{FBA11}.  Also, the Simulated Annealing
method which works so well in panel (b) is highlighted (purple with
shorter dashes) in all panels for comparison.)  Our method is labeled
$U_{opt}$ because numerical optimization over all $\theta \in [0,1]$
has been used to set $\theta$ to the value that maximizes NMI.  Both
the 1000- and 5000-node cases are shown, for small communities (10 to
50 nodes) and large ones (20 to 100).  The $x$-axis is the mixing
parameter $\mu$---the fraction of a node's neighbors outside its
community ({\em not} the expected edge probability $\mu$
of~\eqref{eq:mudef})---and the $y$-axis is a particular version
(cf. the appendix of~\cite{LFK09}) of the Normalized Mutual
Information (NMI) between the computed and the true partition.  In all
cases, our method $U_{opt}$ exhibits the performance characteristic of
the three state-of-the-art methods cited by~\cite{LaFo09}.  The
$U_{opt}$ method has an unfair advantage in optimizing over all
$\theta$: in a deployable algorithm one would need a method for
setting $\theta$.  On the other hand, the purpose of
Figure~\ref{fig:GFcompare} is simply to show that the $\phvw$
computation retains enough information about community structure to
reconstruct high-quality hard-call solutions.  From this perspective,
it is surprising that it does so well, because $U_{opt}$ is based on
(a) the simple utility function of Appendix~\ref{sec:utility}, (b) an
approximation $\ptvw$ to $\pvw$ based only on limited evidence, and
(c) an approximation $\phvw$ to $\ptvw$ based on a heuristic evalution
of the required integral.

\definecolor{uopt}{rgb}{0,0.8,0}
\definecolor{infomap}{rgb}{0,0.3,0.9}
\definecolor{rn}{rgb}{1,0.5,0}
\definecolor{louvain}{rgb}{0.9,0,0.2}


\begin{figure}
\hspace{0.05\columnwidth}
\begin{minipage}[b]{0.4\columnwidth}
  \centering 
    \centerline{\includegraphics[width=\linewidth]{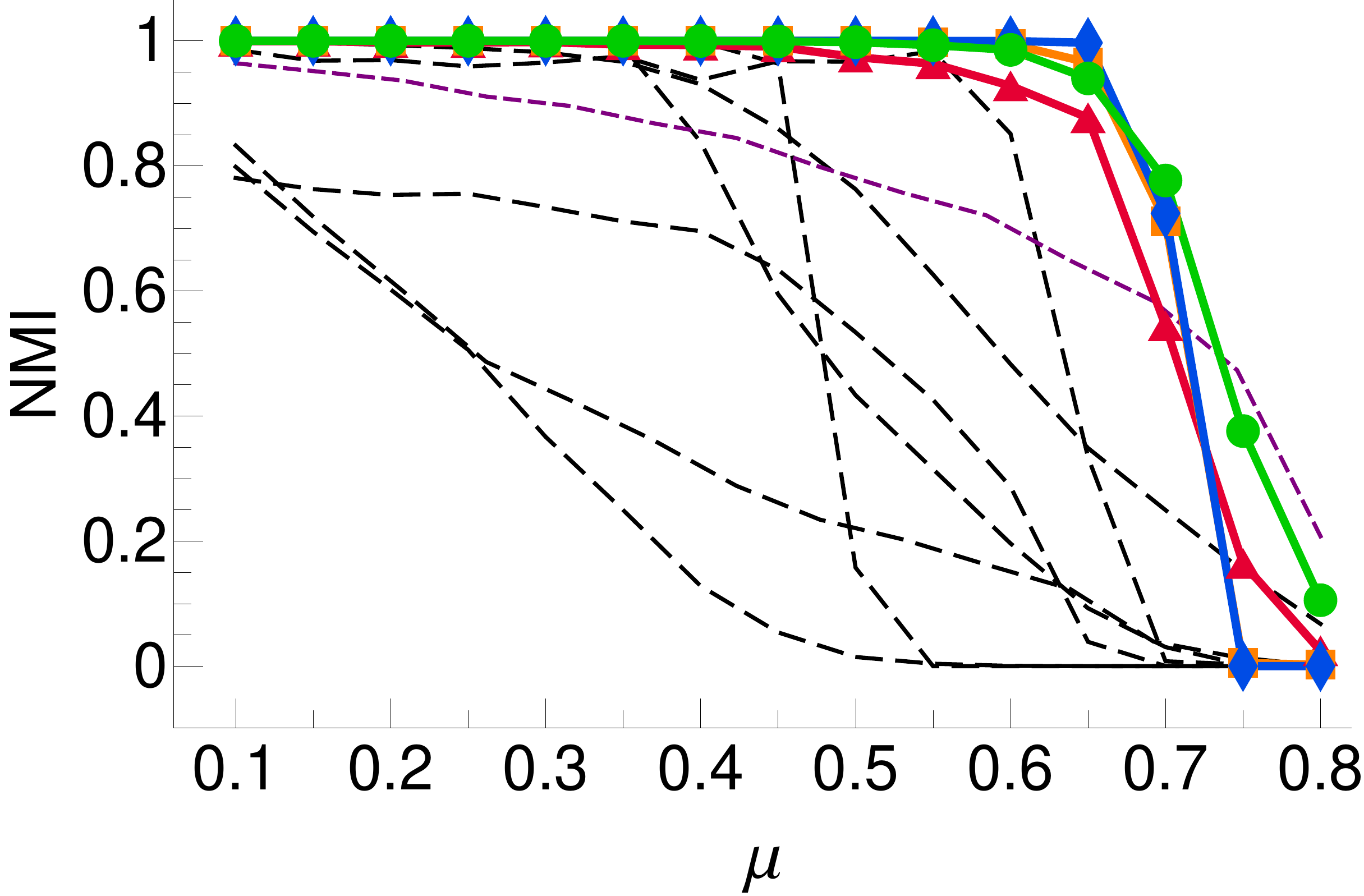}}
  \centerline{(a) 1000 nodes, small communities}\medskip
\end{minipage}
\hfill
\begin{minipage}[b]{0.4\columnwidth}
  \centering
    \centerline{\includegraphics[width=\linewidth]{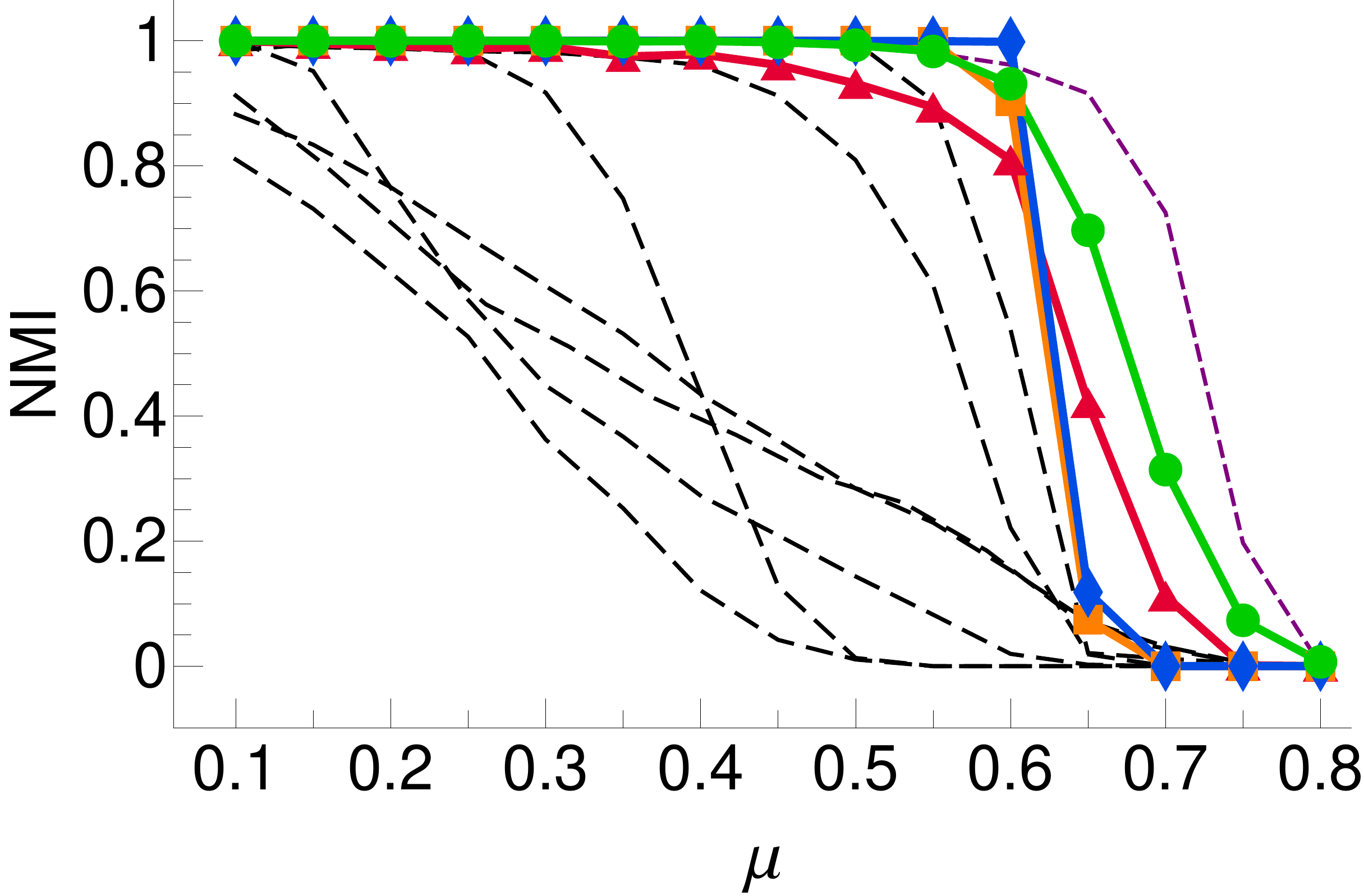}}
  \centerline{(b) 1000 nodes, large communities}\medskip
\end{minipage}
\hspace{0.05\columnwidth}

\hspace{0.05\columnwidth}
\begin{minipage}[b]{0.4\columnwidth}
  \centering
    \centerline{\includegraphics[width=\linewidth]{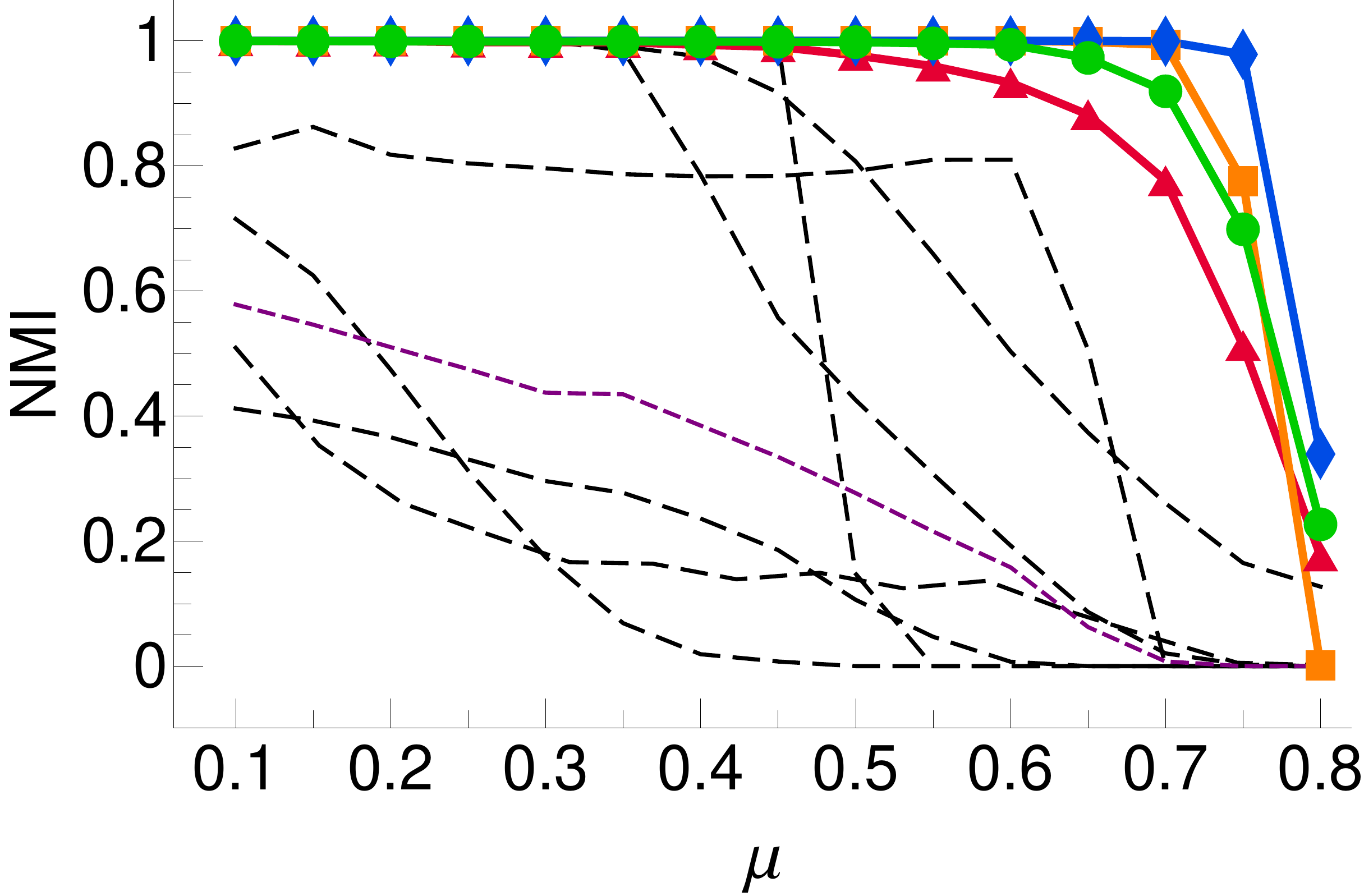}}
  \centerline{(c) 5000 nodes, small communities}\medskip
\end{minipage}
\hfill
\begin{minipage}[b]{0.4\columnwidth}
  \centering
    \centerline{\includegraphics[width=\linewidth]{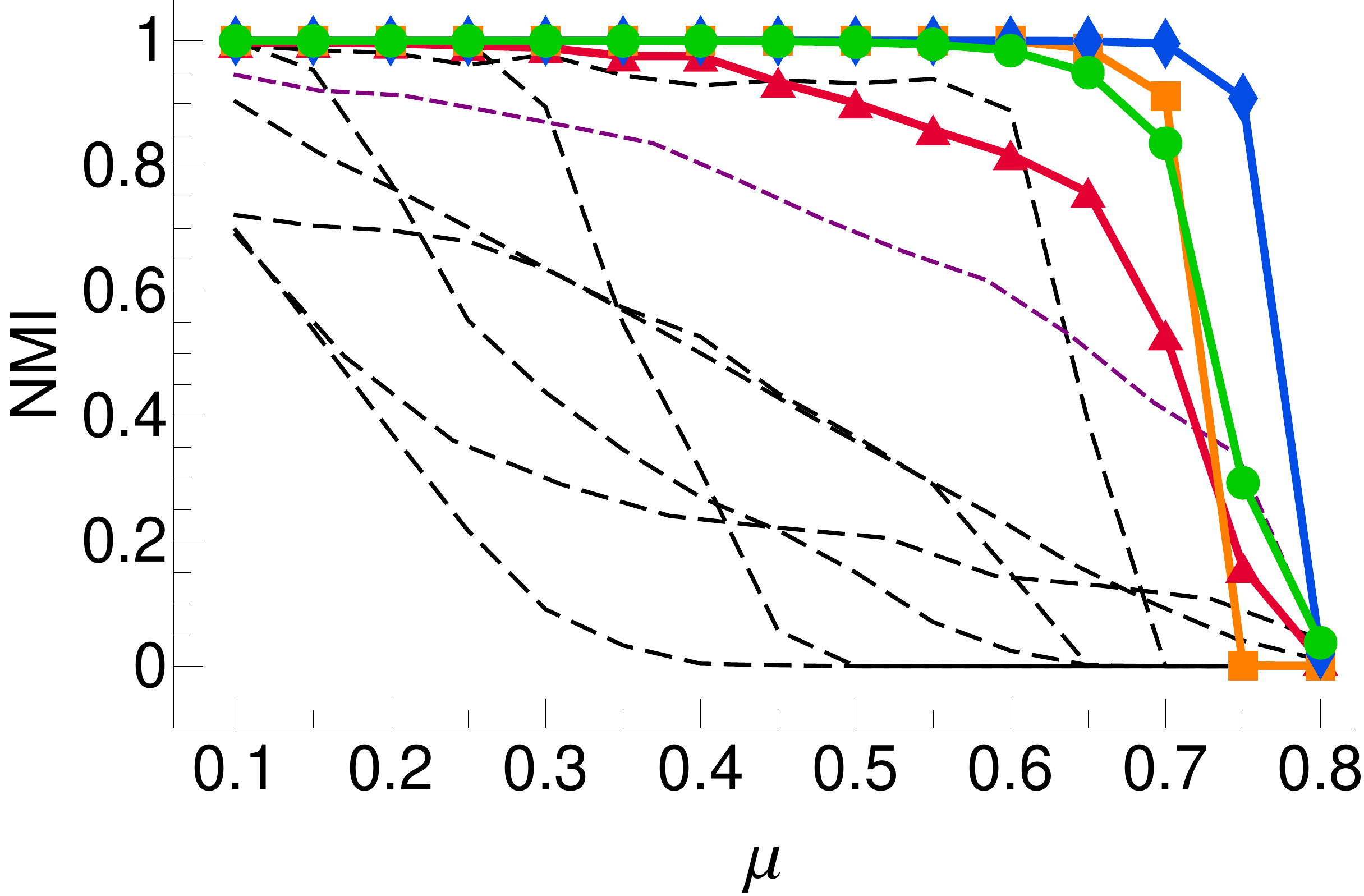}}
  \centerline{(d) 5000 nodes, large communities}\medskip
\end{minipage}
\hspace{0.05\columnwidth}
\caption{Comparison of community detection algorithms:$\;\;$
{\large ${\color{uopt} \jcirc}$}:~$U_{opt}$;
${\color{infomap} \jdiam}$:~Infomap;
${\color{rn} \scriptstyle \jsqua\,}$:~RN;
${\color{louvain} \jtria}$:~Louvain;
- $\!$- $\!$- $\!$:~Other methods.
}
\label{fig:GFcompare}
\end{figure}

\subsection{Efficiency}
\label{sec:efficiency}

The algorithm for computing the $\phvw$ begins with pre-computing
the value of $n_2$ for all pairs of nodes for which $n_2 > 0$, then
creating a cache of $\phvw$ values for triples $(\kappa_{vw},n_1,n_2)$.
For any node pair $\{v,w\}$, the value of $\phvw$ can be
computed by first looking up its value of $n_2$, computing $n_0$ and
$n_1$ from $n$ and the degrees of $v$ and $w$, then looking up the
$\phvw$ value for its triple $(\kappa_{vw},n_1,n_2)$.  Occasionally
the value for this triple must be computed from~\eqref{eq:phvw} and
cached, but the number of such distinct triples is relatively small in
practice.  An optional additional step one can perform is to loop over
all node pairs with non-zero $n_2$ in order to both fill in the
value of $\phvw$ for each triple $(\kappa_{vw},n_1,n_2)$, and count
the number of times each triple occurs.  (Because only the $\{v,w\}$
pairs with $n_2 > 0$ are looped over, some additional bookkeeping is
needed to fill in and provide a count for the $(\kappa_{vw},n_1,0)$ triples
without actually iterating over all $O(n^2)$ node pairs.)  These
values and counts are useful for the statistical analysis of the
$\phvw$ distribution.

We tested the algorithm on five different Facebook networks (gathered
from various universities)~\cite{TKMP11}, and networks generated from
Slashdot~\cite{LLDM09}, Amazon~\cite{LAH07},
LiveJournal~\cite{LLDM09}, and connections between {\tt .edu} domains
(Wb-edu)~\cite{DaHu09}.  Table~\ref{tab:statistics} shows various
relevant network statistics.  The sum of the values $n_2$ for each
node pair is the number of calculations needed to compute the $n_2$
data structure, whereas the number of values of $n_2 > 0$ reflects its
size.  The next column is the number of distinct
$(\kappa_{vw},n_1,n_2)$ triples---this is the number of distinct
$\phvw$ values that must be computed, and the final one is the number
of communities that a randomly chosen instance of the algorithm
Infomap~\cite{RoBe08} found for the dataset.  For the last two rows
the $n_2$ data structure was too large to hold in memory, and the
second step of counting the triples was not performed, nor could
Infomap be run successfully on our desktop.

\begin{table}[h]
  \noindent\centering
  \[ { \small \renewcommand{\arraystretch}{1.2}
\begin{array}{|l|r|r|r|r|r|r|}\hline
\multicolumn{1}{|c|}{\text{Dataset}} &
\multicolumn{1}{c|}{\;\text{Nodes}\;} &
\multicolumn{1}{c|}{\;\text{Edges}\;} &
\multicolumn{1}{c|}{\;\sum n_2\;} &
\multicolumn{1}{c|}{\;\#(n_2 > 0)\;} &
\multicolumn{1}{c|}{\;\;\text{Triples}\;} &
\multicolumn{1}{c|}{\;\text{Groups}\;} \\\hline
\text{Caltech}		&769			&16,\!656		&1,\!231,\!412			&186,\!722			&14,\!120	&19		\\\hline
\text{Princeton}	&6,\!596		&293,\!320		&46,\!139,\!701			&8,\!776,\!074			&83,\!004	&51		\\\hline
\text{Georgetown}	&9,\!414		&425,\!638		&67,\!751,\!053			&15,\!616,\!610			&113,\!722	&90		\\\hline
\text{Oklahoma}		&17,\!425		&892,\!528		&194,\!235,\!901		&47,\!202,\!925			&239,\!162	&233		\\\hline
\text{UNC}		&18,\!163		&766,\!800		&140,\!796,\!299		&47,\!576,\!619			&191,\!482	&167		\\\hline
\text{Slashdot}		&82,\!168		&504,\!230		&74,\!983,\!589			&49,\!450,\!449			&104,\!330	&5,\!209	\\\hline
\text{Amazon}		&262,\!111		&899,\!792		&9,\!120,\!350			&6,\!434,\!638			&7,\!178	&\;12,\!851\;	\\\hline
\text{LiveJournal}	&4,\!847,\!571		&42,\!851,\!237		&7,\!269,\!503,\!753		&4,\!193,\!393,\!006		&		&		\\\hline
\text{Wb-edu}		&\;9,\!450,\!404\;	&\;46,\!236,\!105\;	&\;12,\!203,\!737,\!639\;	&\;4,\!232,\!928,\!806\;	&		&		\\\hline
\end{array}
} \] \caption{Network datasets} \label{tab:statistics}
\end{table}

Table~\ref{tab:timing} contains timing results based on a Dell desktop
with 8GB of RAM, and eight 2.5GHz processors.  The first column is the
number of seconds it took the version of Infomap described
in~\cite{RoBe08} to run.  This code is in C++, runs single-threaded,
includes a small amount of overhead for reading the network, and uses
the Infomap default setting of picking the best result from ten
individual Infomap trial partitions.  The next four columns compare
methods of using our Java code to compute $n_2$.  The first two are
single-threaded, and the other two use all eight cores.  The columns
labeled $n_2 \rightarrow \emptyset$ are the timing for the computation
only: results are nulled out immediately after computing them.  These
columns are included for two reasons.  First, they show that the
computation itself displays good parallelization: the speedup is
generally higher than 6.5 for eight processors.  Second, the
computation itself for the two larger datasets is quite fast (just
under three minutes on eight cores), but the algorithm is currently
designed only to maintain all results in RAM, and these datasets are
too large for this.  The last two columns are the timing results for
explicitly filling in the $\phvw$ information ahead of time and
providing the counts required for statistical analysis.

As Infomap is one of the fastest community detection algorithms
available, these results are quite impressive.  Comparing the first
two columns of timing data, we find the speed-up ranges up to 107 and
862 times as fast as Infomap for the two largest networks on which we
ran Infomap, respectively.  The relative performance falls off rapidly
for denser networks, but even in these cases or method performed
roughly 10 times as fast as Infomap (i.e., as fast as an individual
Infomap run).  Computing the counts for $\phvw$ statistics increases
the run time, but only by a constant factor (of 2 to 3).  It must be
emphasized that the timing results are for producing a very different
kind of output than Infomap does.  However, the usual method for
estimating $\pvw$~\cite{ReBo06} is many times slower than producing
partitions, rather than many times faster.

\begin{table}[h]
  \noindent\centering
  \[ { \small \renewcommand{\arraystretch}{1.1}
\begin{array}{|l|D{.}{.}{1}|D{.}{.}{2}|D{.}{.}{2}|D{.}{.}{2}|D{.}{.}{2}|D{.}{.}{2}|D{.}{.}{2}|}\hline
\multicolumn{1}{|c|}{\multirow{2}{*}{Dataset}} &
\multicolumn{1}{c|}{\;\text{Infomap}\;} &
\multicolumn{1}{c|}{n_2} &
\multicolumn{1}{c|}{\;n_2 \rightarrow \emptyset\;} &
\multicolumn{1}{c|}{n_2} &
\multicolumn{1}{c|}{\;n_2 \rightarrow \emptyset\;} &
\multicolumn{1}{c|}{\;\text{Count}\;} &
\multicolumn{1}{c|}{\;\text{Count}\;} \\
\multicolumn{1}{|c|}{} &
\multicolumn{1}{c|}{\;10 \;\text{trials}\;} &
\multicolumn{1}{c|}{\;1 \;\text{proc}\;} &
\multicolumn{1}{c|}{\;1 \;\text{proc}\;} &
\multicolumn{1}{c|}{\;8 \;\text{proc}\;} &
\multicolumn{1}{c|}{\;8 \;\text{proc}\;} &
\multicolumn{1}{c|}{\;1 \;\text{proc}\;} &
\multicolumn{1}{c|}{\;8 \;\text{proc}\;} \\\hline
\text{Caltech}		&1.0	&0.11	&0.09	&0.02	&0.02	&0.09	&0.03	\\\hline
\text{Princeton}	&45.4	&4.5	&3.9	&0.67	&0.55	&3.9	&0.70	\\\hline
\text{Georgetown}	&77.5	&7.2	&6.4	&1.2	&0.93	&7.0	&1.2	\\\hline
\text{Oklahoma}		&310.8	&23.5	&19.4	&4.2	&2.8	&33.7	&4.6	\\\hline
\text{UNC}		&446.6	&19.2	&16.3	&3.6	&2.5	&22.9	&4.0	\\\hline
\text{Slashdot}		&1553.4	&14.5	&12.6	&3.0	&1.9	&21.8	&3.6	\\\hline
\text{Amazon}		&2075.8	&2.4	&2.2	&0.59	&0.44	&3.2	&0.69	\\\hline
\text{LiveJournal}	&	&	&1246.2	&	&172.9	&	&	\\\hline
\text{Wb-edu}		&	&	&1243.0	&	&174.6	&	&	\\\hline
\end{array}
} \] \caption{Timing results (in seconds)} \label{tab:timing}
\end{table}

\subsection{Uses}
\label{sec:uses}

Visualizing the $\phvw$ matrix provides insight into the hierarchical
community structure of a network~\cite{ReBo06}.  To do so requires an
appropriate ordering on the nodes: one that places nodes nearby when
they are in the same small community, and small communities nearby
when they are in the same larger community.  Therefore, it is useful
to have a dendrogram of hierarchical community clustering.  There are
standard routines for producing such dendrograms, provided one has a
distance metric between points~\cite{JaDu88}.  The community
non-co-membership probabilities $p^{\{v\}\{w\}} = 1 - \pvw$ may be
used for this purpose.  They satisfy the triangle inequality
$p^{\{v\}\{w\}} \le p^{\{v\}\{x\}} + p^{\{w\}\{x\}}$ (although
distinct nodes have distance 0 between them if they are known to be in
the same community).  The approximation~\eqref{eq:phvw} does not obey
this triangle inequality, however, and~\cite{FeBu10} indicates that
this can cause problems in certain contexts.  An approximation of
$\pvw$ which does obey it is a topic for future research.  In the
meantime, we will rely on this version, which tends to obey it for
most node triples.

\begin{figure}
\begin{minipage}[b]{.61\linewidth}
  \centering 
    \centerline{\includegraphics[height=44mm]{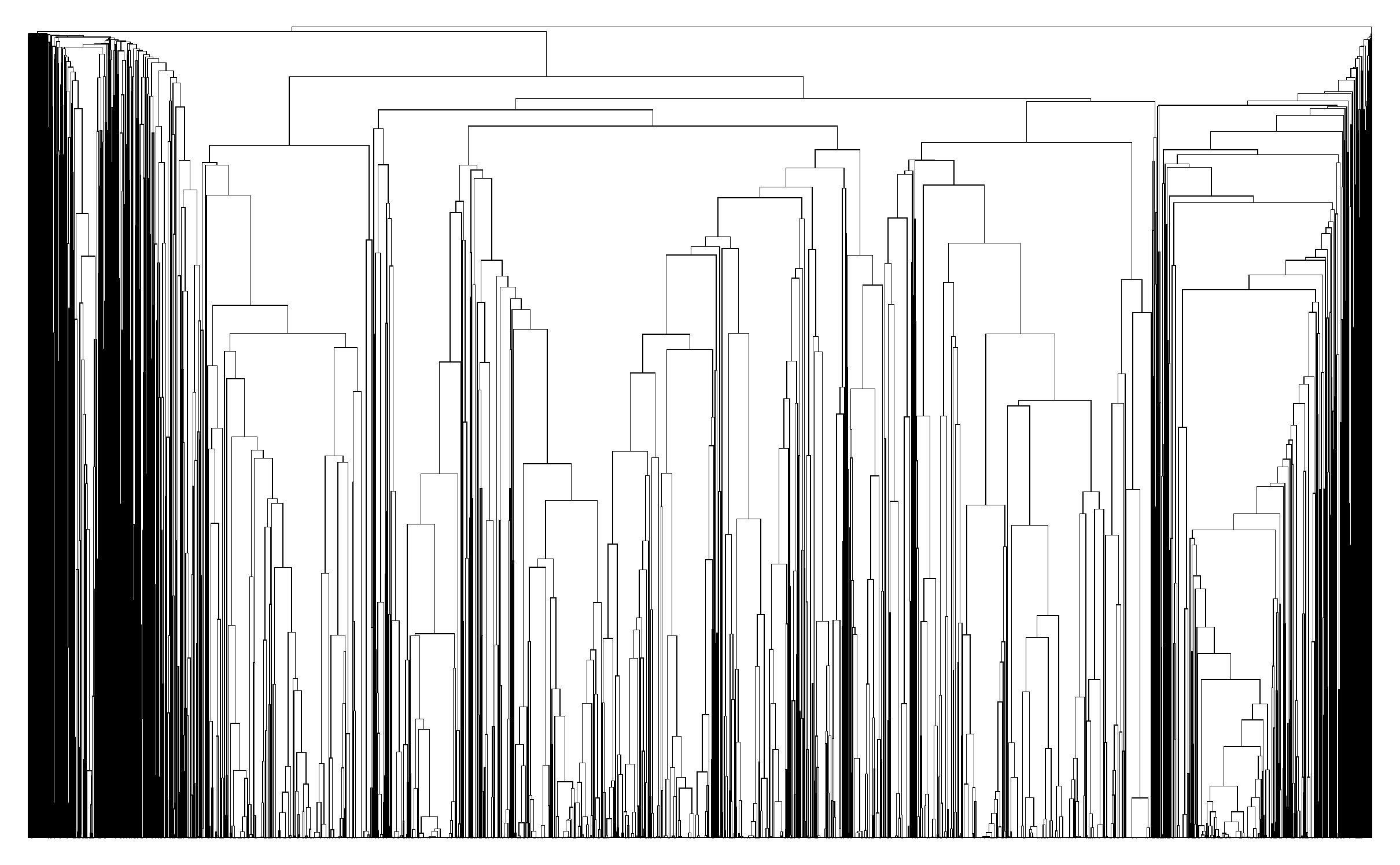}}
  \centerline{(a) Dendrogram for node ordering}\medskip
\end{minipage}
\hfill
\begin{minipage}[b]{0.39\linewidth}
    \centerline{\includegraphics[height=44mm]{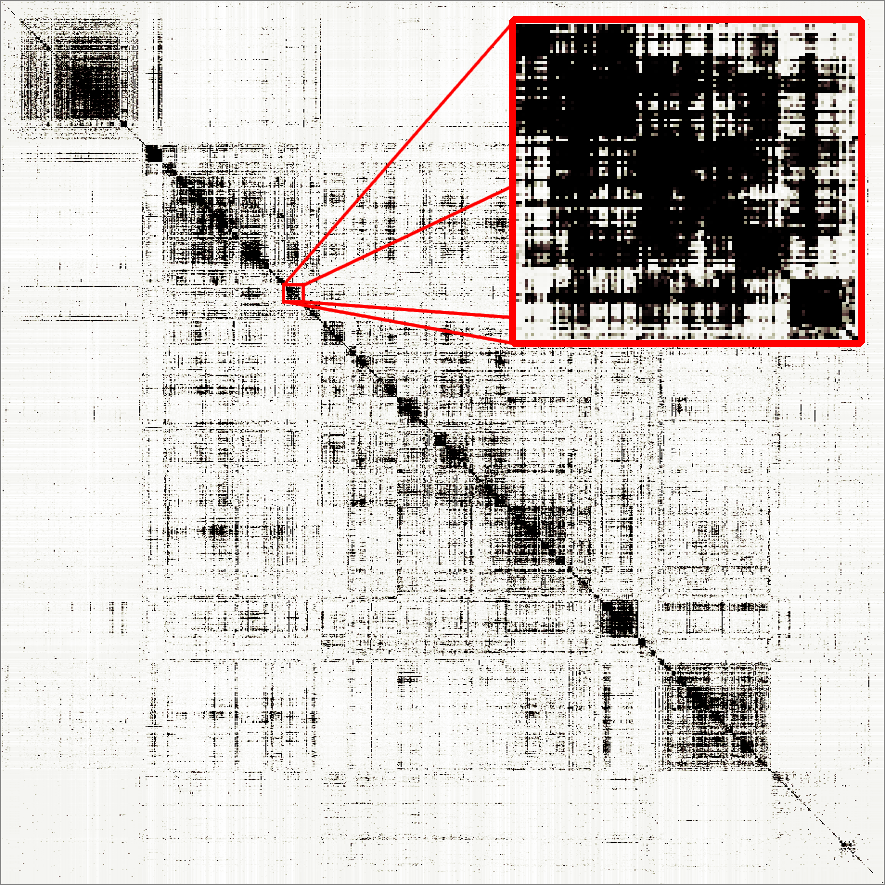}}
  \centerline{(b) $\phvw$ matrix}\medskip
\end{minipage}
\caption{Visualization of community structure for Princeton Facebook network}
\label{fig:pvwplots}
\end{figure}

To order the nodes of a graph we use this (approximate) distance
metric $p^{\{v\}\{w\}}$ with a hierarchical clustering scheme that
defines cluster distance as the average distance between nodes.
The output of this is a binary tree (ties having been broken
arbitrarily), so the order of the clusters at each branch point must
still be determined.  This is done by starting at the root of the tree
and testing which ordering at each branch point yields a smaller
average distance to its neighboring clusters on the right and left.
Figure~\ref{fig:pvwplots} shows the resulting dendrogram for the
Princeton Facebook network (cf. Table~\ref{tab:statistics}), and the
corresponding $\phvw$ matrix.  (This method was used to generate the
ordering in Figure~\ref{fig:enron} as well.)

\begin{figure}
  \centering
    \centerline{\includegraphics[width=0.6\columnwidth]{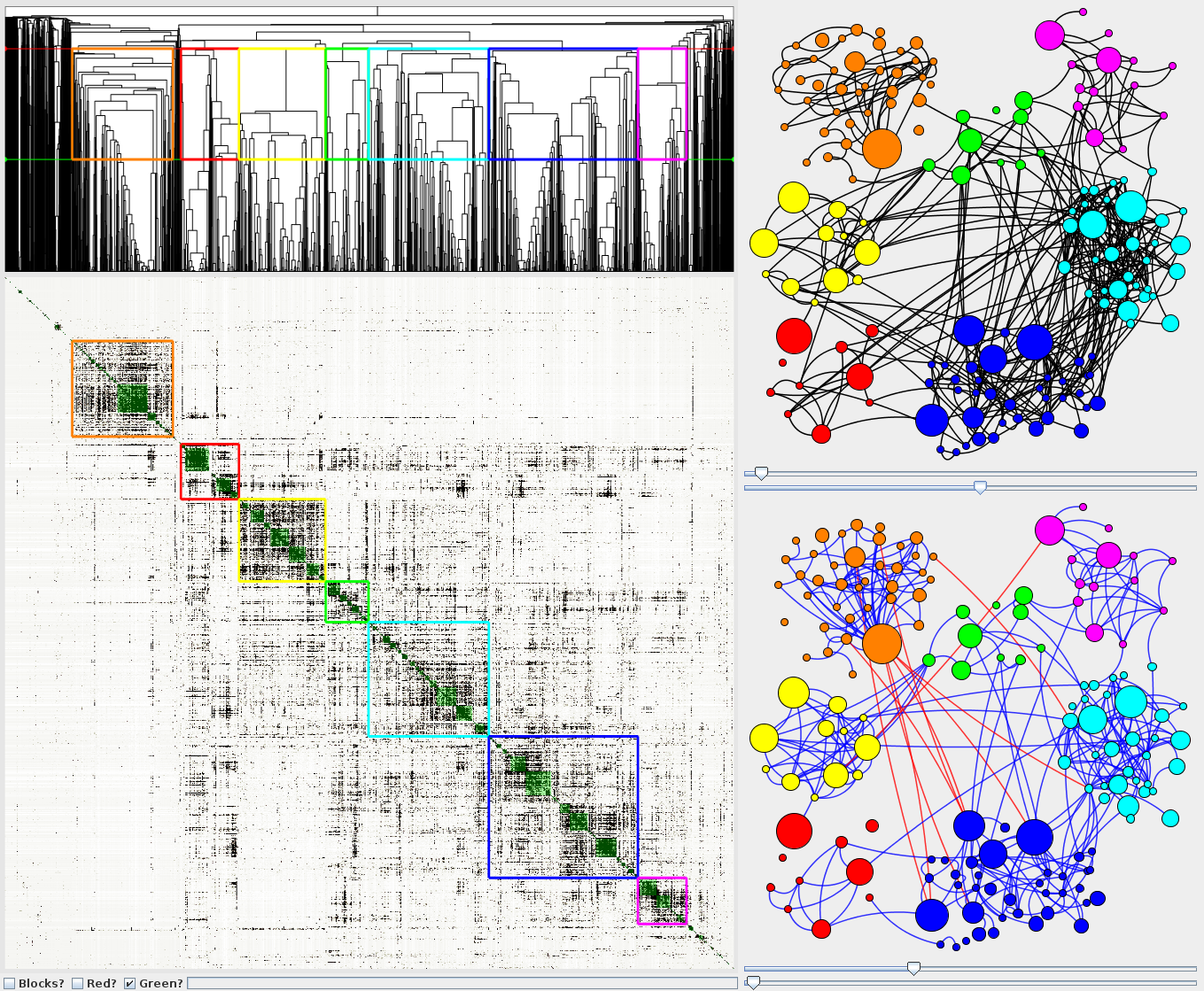}}
%
\caption{Screen shot of IGNITE network probability visualization tool
  for Georgetown Facebook network}
\label{fig:screenshot}
\end{figure}

Visualizations like Figure~\ref{fig:pvwplots} are useful for network
analysis.  We have combined them other visualizations in the code
IGNITE (Inter-Group Network Inference and Tracking Engine).
Figure~\ref{fig:screenshot} uses IGNITE to the Georgetown Facebook
network (cf. Table~\ref{tab:statistics}).  The dendrogram on which the
ordering for the $\phvw$ matrix is based is shown in the upper left
panel.  Two levels in this dendrogram have been selected: the lower
level is used to coarse-grain the network by merging communities of
nodes together into meta-nodes; the upper level is used to determine
which sets of meta-nodes to consider communities.  The selection of
these levels is reflected in the $\phvw$ matrix panel below.  The
meta-nodes are indicated by translucent green squares, and communities
of nodes are outlined in different colors (corresponding to similar
outlines in the dendrogram).  The meta-nodes and communities are then
displayed in panels on the right: the upper panel corresponding to a
coarse-grained version of the original network; the lower, to a
variant where the edges have been replaced with averaged $\phvw$
values between meta-nodes.  The sizes of the meta-nodes indicates how
many true nodes they contain.  In the lower right panel, blue lines
indicate average $\phvw$ above a certain threshold, and red below
another (much lower) threshold, as in Figure~\ref{fig:pvwkarate}.

\section{Community Tracking:  Exact Equations}
\label{sec:dynamicExact}

The study of dynamic networks is much less developed than its static
counterpart. There is substantial work on processes evolving on
networks.  For example, see~\cite{Dan06} for a discussion of the complex
dynamical systems which arise in economics and traffic engineering,
along with mathematically rigorous results about their equilibria.
Diffusion equations on networks have particularly elegant properties.
They are governed by the {\em Laplacian} matrix of a graph, the
discrete analog of continuum Laplacian operator, and are therefore an
important topic in spectral graph theory~\cite{NLCK05}.  These may be
generalized to reaction--diffusion equations and used to model the
spread of disease~\cite{CPV07}, but the more common model in network
epidemiology is the SIR model~\cite{KeMc27}.  Such models have been
extended to model the spread of rumors~\cite{BBR06},
obesity~\cite{ChFo07}, and innovations~\cite{Val96}.

The term ``dynamic networks'' implies that the networks themselves are
evolving in time, however.  Stokman and Doreian edited several
influential special editions of the Journal of Mathematical Sociology
on network evolution, the first of which was in 1996 and published in
book form as~\cite{DoSt97}. This work illustrated how macroscopic
behavior of network evolution arises from local governing
laws. Snijders emphasizes~\cite{SBS10} the benefits of casting the
dynamic network problem in the continuous-time, Markov process
framework first proposed by Leenders~\cite{Lee95}.  In particular,
there is a small body of literature on communities
evolving in dynamic networks. Much of this work is summarized in 3 1/2
pages of Fortunato's 100-page review of group
finding~\cite{For10}. The field begins with the 2004 work of Hopcroft
et al.~\cite{HKKS04}, which studied the persistence of robust
communities in the NEC CiteSeer database.  The most prominent
publication is the 2007 work of Palla et al.~\cite{PBV07}, which
analyzed the evolution of overlapping groups in cell phone and
co-authorship data and presented a method for tracking communities
based on the clique percolation method used in CFinder~\cite{PDFV05}.
The various researchers in the community have come to agree on the key
fundamental events of community evolution: birth/death,
expansion/contraction, and merging/splitting~\cite{GDC10}.
Berger-Wolf and colleagues propose an optimality criterion for
assigning time-evolving community structure to a sequence of network
snapshots, prove that it is NP-hard to find the optimal structure, and
develop various approximation techniques~\cite{TBK07,TaBe09,SBG10}.

Most of the work on community tracking considers discrete network
snapshots and attempts to match up the community structure at
different time steps.  From the perspective of the data fusion
community, such an approach to tracking may seem {\it ad hoc}: one could
argue that (a) the ``matching up'' criteria are necessarily heuristic,
and (b) one gets only a single best solution with no indication of the
uncertainty.  In contrast, the tracking work in data fusion is based
on formal evolution and measurement models for the full probability
distribution over some state space, followed by principled
approximations~\cite{Jaz70}.  A sensible response to this critique,
however, is that the state space in the community tracking problem is
so much larger that data-fusion-style tracking techniques do not
apply.  The truth is perhaps somewhere in between: it is, in fact,
possible to derive a formal Bayesian filter for the community tracking
problem and to produce tractable approximations to it.  Indeed, this
is the topic of the remainder of this paper.  On the other hand, the
filter derived is more a proof of concept than an algorithm ready to
supplant the more informal methods.  It may be that the formal
approach we present here can be developed into a true, practical
``Kalman filter for networks.''  On the other hand, it may be that
concerns about uncertainty management can be addressed without appeal
to a formal model.  For example, Rosvall and Bergstrom have devised a
re-sampling technique to estimate the degree to which the data support
the various assignments of nodes to time-evolving
communities~\cite{RoBe10}.  Similarly, the work of Fenn et al. tracks
the evolution of groups by gathering evidence that each node belongs
to one of a number of known groups~\cite{FPMW09}, thus providing
output similar to one version of the method outlined below.

We model community and graph evolution as a continuous-time Markov
process~\cite{CoMi65}, $\{(\Phi_t,\Kappa_t): t \ge 0\}$, the
continuous-time analog of a Markov chain.  The continuous-time
approach is more general (in that it can always be sampled at discrete
times to produce a Markov chain) and is simpler to work with due to
the sparsity of the matrices involved.  We will not explicitly
indicate any dependence on structural parameters $\params$, because we
will not integrate these out as was done in the static case.  In this
section we will use $\Kappa_{[0,t]} \doteq \{\Kappa_{t'} : 0 \le t' \le
t\}$ to denote the time-history of the network process up through time
$t$, and $\Kappa_{[0,t)} \doteq \{\Kappa_{t'} : 0 \le t' < t\}$ for the
history not including the current time $t$ (with similar definitions
for $\Phi_{[0,t]}$ and $\Phi_{[0,t)}$).  The purpose of this section
is to derive a Bayesian filter $\pr(\Phi_t | \Kappa_{[0,t]})$: i.e.,
assuming some initial distribution $\pr(\Phi_0)$ is given,
Section~\ref{sec:full} derives the expressions for evolving the
distribution of the community structure to time $t$, given all network
evidence up through time $t$.  In the community detection case, the
next step was to approximate marginals of the full distribution using
limited graph evidence.  In the tracking case, however, it is possible
to obtain exact formulas for the marginals: this is done in
Section~\ref{sec:marginalization}.  These formulas, though exact, are
not {\em closed}, however: the approximations required to close them
are discussed in Section~\ref{sec:dynamicApprox}.

\subsection{Evolution of the full distribution}
\label{sec:full}

A {\em dynamic stochastic blockmodel} $\cH(n,A,\bB)$ may be defined
analogously to the static version $H(n,\bp,\bQ)$ introduced in
Section~\ref{sec:block}.  Whereas $H(n,\bp,\bQ)$ defines a pair of
random variables $\Phi$ and $\Kappa$, $\cH(n,A,\bB)$ defines a pair of
stochastic processes $\{\Phi_t: t \ge 0\}$ and $\{\Kappa_t: t \ge
0\}$.  The joint process $\{(\Phi_t,\Kappa_t): t \ge 0\}$ will be
modeled as a {\em continuous-time Markov process}.  The parameter $A$
is an $m \times m$ matrix, and $\bB$ is a collection of $r \times r$
matrices $B_{ij}$ for $1 \le i \le j \le m$ (and for convenience we
define $B_{ji} = B_{ij}$ for $j > i$).  Just as $\bp$ and the
$\bq_{ij}$ were required to be stochastic vectors (i.e., vectors with
non-negative entries that sum to one) in Section~\ref{sec:block}, so
the matrices $A$ and $B_{ij}$ are required to be {\em transition rate
matrices}: i.e., they must have non-negative off-diagonal entries, and
their columns must sum to zero.  The entry $a_{ij}$ of $A$ defines the
transition rate of a node in group $j$ switching to group $i$: i.e.,
the probability of a node in group $j$ being in group $i$ after an
infinitesimal time $\Delta t$ is $\delta_{ij} + a_{ij} \, \Delta t +
O((\Delta t)^2)$.  Similarly, the entry $b_{ij,kl}$ of $B_{ij}$
defines the rate that an edge connecting nodes in groups $i$ and $j$
transitions from edge type $l$ to type $k$.

\begin{figure}
  \centering
    \centerline{\includegraphics[width=0.55\columnwidth]{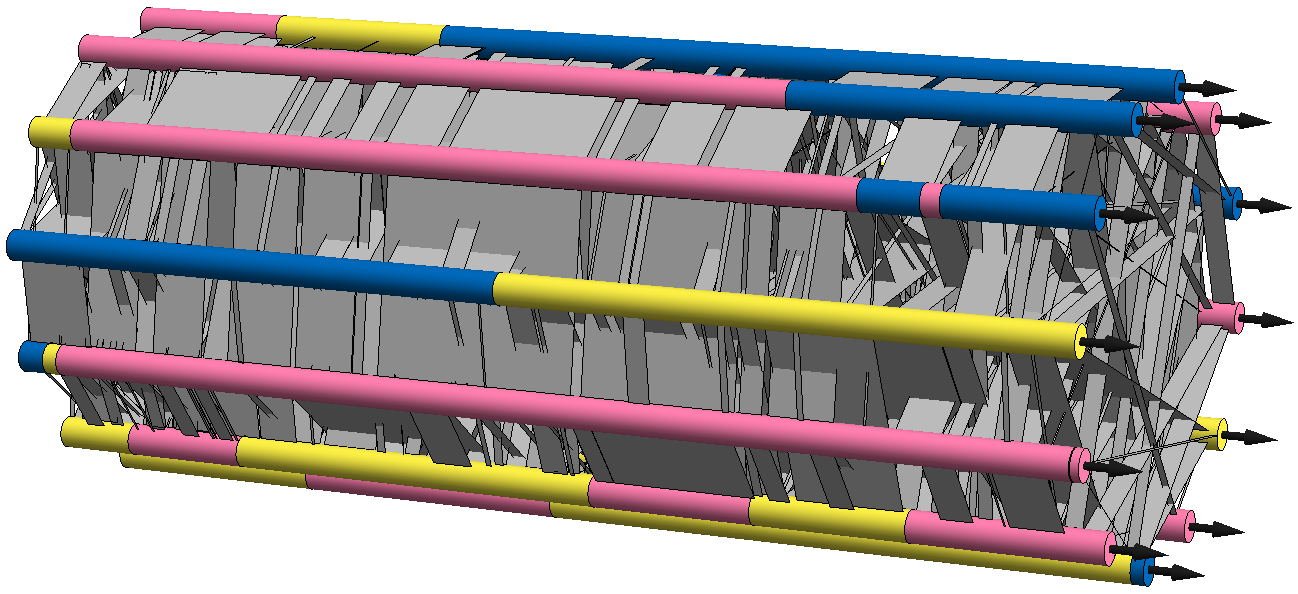}}
%
\caption{An instance of $\cH(12,3,0.5,16,4,2,18)$}
\label{fig:rocket}
\end{figure}

We may define a {\em dynamic planted partition model}
$\cH(n,m,a,\lambda_I,\mu_I,\lambda_O,\mu_O)$ as a special case of
$\cH(n,A,\bB)$.  As in Section~\ref{sec:planted}, this cases is
obtained by requiring that $A$ and $\bB$ be invariant under
permutations of $[m]$ and using only $r = 2$ edge types (``off''
($k=0$) and ``on'' ($k=1$)).  In this case, the transition rate matrix
$A$ reduces to a single rate $a$ at which nodes jump between
communities, while the collection $\bB$ of transition rate matrices
reduces to four rate parameters:  $\lambda_I$, the rate at which edges
turn on for pairs of nodes in the same community; $\mu_I$, the rate at
which edges turn off for pairs of nodes in the same community; and
$\lambda_O$ and $\mu_O$, the corresponding rates for pairs of nodes in
different communities.  Figure~\ref{fig:rocket} depicts and instance of this
model with $n = 12$ nodes and $m = 3$ communities with rate parameters
$a = 0.5$, $\lambda_I = 16$, $\mu_I = 4$, $\lambda_O = 2$, and $\mu_O
= 18$ for $t = 0$ to 5.

If two independent random variables $X$ and $Y$ have respective
probabilities $x_i$ and $y_j$ for their various outcomes $i$ and $j$,
then the joint random variable $Z = (X,Y)$ has outcomes indexed by
$(i,j)$ with probabilities $z_{(i,j)} = x_i y_j$.  The analog of this
for Markov processes is expressed by the {\em Kronecker
sum}~\cite{Ort87}.  I.e., Suppose two independent Markov processes
$\{X(t) : t \ge 0\}$ and $\{Y(t) : t \ge 0\}$ have respective
probabilities $x_i(t)$ and $y_j(t)$ for their various outcomes $i$ and
$j$ at time $t$, and that these probabilities are governed by
$\dot{\mathbf{x}} = A \mathbf{x}$ and $\dot{\mathbf{y}} = B
\mathbf{y}$, respectively (where $\mathbf{x}(t)$ collects all the
$x_i(t)$, and $\mathbf{y}(t)$, the $y_j(t)$).  Then the joint Markov
process $\{Z(t) : t \ge 0\}$, where $Z(t) = (X(t),Y(t))$, has outcomes
indexed by $(i,j)$ with probabilities $z_{(i,j)}$ that are governed by
$\dot{\mathbf{z}} = (A \oplus B) \mathbf{z}$ (where $\mathbf{z}(t)$
collects all the $z_{(i,j)}$).  The Kronecker sum $A \oplus B$ is
defined by
\begin{equation}
(A \oplus B)_{(i,j),(i',j')} \doteq a_{ii'} \delta_{jj'} + \delta_{ii'} b_{jj'}.
\end{equation}
The interpretation of this is that in an infinitesimal time the Markov
process $X_t$ may transition from $i$ to another state $i'$, or $Y_t$
may transition from $j$ to $j'$, but for both to change simultaneously
is infinitely less likely than for only one to change.

The derivation of the Bayesian filter for $\pr(\Phi_t |
\Kappa_{[0,t]})$ follows the same six steps as the static derivation
in Section~\ref{sec:block}.  Indeed, the purpose of including the
six-step derivation in Section~\ref{sec:block} was to make the
following derivation easier to follow by analogy.

\uhead{Step 1.}  Let $\bp^v(t) \in \mathbb{R}^m$ denote the vector of
probabilities $p^v_i(t)$ that a single node $v$ is in community $i$ at
time $t$:  i.e., $p^v_i(t) \doteq \pr(\Phi_t(v) = i)$.  These probabilities
are governed by the transition rate matrix $A$.  Therefore
\begin{equation}
\frac{d\bp^v}{dt} = A \bp^v, \quad \text{which has the solution} \;\;
\bp^v(t') = e^{A(t'-t) \bp^v(t)}.
\end{equation}

\uhead{Step 2.}  Let $\pvp \in \mathbb{R}^{m^n}$ denote the
vector of probabilities $\pcp_\phi(t)$ that the communities of all $n$
nodes are specified by the assignment $\phi$ at time $t$:  i.e.,
$\pcp_\phi(t) \doteq \pr(\Phi_t = \phi)$.  The transition rate matrix for
this joint process on all nodes is the Kronecker sum of the (identical)
transition rate matrices for each node:
\begin{equation}
\frac{d\pvp}{dt} = \tmp \pvp, \quad \text{where} \;\;
\tmp = \bigoplus_{v=1}^n A.
\end{equation}
The components $\tcp_{\phi' \phi}$ of $\tmp$ may be expressed as
\begin{equation}
\tcp_{\phi' \phi} = \begin{cases} \displaystyle
\sum_{v=1}^n a_{\phi(v) \phi(v)},\quad &\text{if}\;\phi'=\phi,\\
a_{\phi'(v^*) \phi(v^*)}, &\text{if}\; \phi'(v) = \phi(v) \;\text{for all}
\; v \neq v^*,\\
0, &\text{otherwise}.
\end{cases}
\end{equation}

\uhead{Step 3.}  Let $\bq^e(t) \in \mathbb{R}^N$ (where $N =
n(n-1)/2$) denote the vector of probabilities $q^e_k(t)$ that a single
edge $e$ has type $k$ at time $t$ given the current communities of its
endpoints: i.e., $q^e_k(t) \doteq \pr(\Kappa_t(e) = k |
\Phi_t(v),\Phi_t(w))$.  These probabilities are governed by the
transition rate matrix $B_{ij}$, where $i = \phi_t(v)$ and $j =
\phi_t(w)$ are the current communities of $v$ and $w$:
\begin{equation}
\frac{d\bq^e}{dt} = B_{\phi_t(v)\phi_t(w)} \bq^e. \label{eq:qe}
\end{equation}
The matrix $B_{\phi_t(v)\phi_t(w)}$ is piecewise constant in time, so
the solution to~\eqref{eq:qe} is a (continuous) piecewise exponential
function.

\uhead{Step 4.}  Let $\pvk \in \mathbb{R}^{r^N}$ denote the
vector of probabilities $\pck_\kappa(t)$ that the graph is $\kappa$ at
time $t$ given the current communities of all nodes:  i.e.,
$\pck_\kappa(t) \doteq \pr(\Kappa_t = \kappa | \Phi_t)$.  The
transition rate matrix for this joint process on all edges is the
Kronecker sum of the transition rate matrices for each edge:
\begin{equation}
\frac{d\pvk}{dt} = \tmk_{\phi_t} \pvk, \quad \text{where} \;\;
\tmk_\phi = \bigoplus_{e \in \eset} B_{\phi(e_1)\phi(e_2)}.
\end{equation}
The components $\tck_{\phi,\kappa'\kappa}$ of $\tmk_\phi$ may be expressed as
\begin{equation}
\tck_{\phi,\kappa'\kappa} = \begin{cases} \displaystyle
\sum_{e \in \eset} b_{\phi(e_1) \phi(e_2),\kappa(e)\kappa(e)},\quad &\text{if}\;\kappa'=\kappa,\\
b_{\phi(e_1^*) \phi(e_2^*),\kappa'(e^*)\kappa(e^*)}, &\text{if}\; \kappa'(e) = \kappa(e) \;\text{for all}
\; e \neq e^*,\\
0, &\text{otherwise}.
\end{cases}
\end{equation}

\uhead{Step 5.}  Let $\pvkp \in \mathbb{R}^{m^n r^N}$ denote the
vector of probabilities $\pckp_{(\phi,\kappa)}$ that the community
assignment is $\phi$ and the graph is $\kappa$ at time $t$:  i.e.,
$\pckp_{(\phi,\kappa)}(t) \doteq \pr(\Phi_t = \phi, \Kappa_t = \kappa)$.
The transition rate matrix for this process is not quite a Kronecker
sum due to the dependence of $\tmk_\phi$ on $\phi$---it loses
the various nice properties that Kronecker sums have, but the formula
is quite similar:
\begin{equation}
\frac{d\pvkp}{dt} = \tmkp \pvkp, \quad \text{where} \;\;
\tckp_{(\phi',\kappa')(\phi,\kappa)} = \tcp_{\phi'\phi}
\delta_{\kappa'\kappa} + \delta_{\phi'\phi}
\tck_{\phi,\kappa'\kappa}. \label{eq:dpvkp}
\end{equation}
A Bayesian filter has a prediction step (which applies while the graph
data remains constant) and an update step (which applies when the
graph data changes).  Therefore, we need to decompose~\eqref{eq:dpvkp}
into a component which is zero while the graph is constant and a
component which is zero when the graph changes.  The required
decomposition uses slightly modified
matrices $\tmp'_\kappa$ and $\tmk'_\phi$:
\begin{align}
\tckp_{(\phi',\kappa')(\phi,\kappa)} &= \tcp'_{\kappa, \phi'\phi}
\delta_{\kappa'\kappa} + \delta_{\phi'\phi}
\tck'_{\phi,\kappa'\kappa}, \quad \text{where} \\
\tcp'_{\kappa, \phi'\phi} &= \tcp_{\phi'\phi} + \delta_{\phi'\phi}
\tck_{\phi,\kappa\kappa}, \quad \text{and} \quad
\tck'_{\phi,\kappa'\kappa} = \tck_{\phi,\kappa'\kappa} -
\delta_{\kappa'\kappa} \tck_{\phi,\kappa\kappa}.
\end{align}
The components $\tcp'_{\kappa, \phi'\phi}$ of $\tmp'_\kappa$ may be expressed as
\begin{equation}
\tcp'_{\kappa, \phi' \phi} = \begin{cases} \displaystyle
\sum_{v=1}^n a_{\phi(v) \phi(v)} + \!\!\sum_{e \in \eset} b_{\phi(e_1) \phi(e_2),\kappa(e)\kappa(e)},\quad &\text{if}\;\phi'=\phi,\\
a_{\phi'(v^*) \phi(v^*)}, &\text{if}\; \phi'(v) = \phi(v) \;\text{for all}
\; v \neq v^*,\\
0, &\text{otherwise}.
\end{cases}
\end{equation}
The components $\tck'_{\phi,\kappa'\kappa}$ of $\tmk'_\phi$ may be expressed as
\begin{equation}
\tck'_{\phi,\kappa'\kappa} = \begin{cases} \displaystyle
b_{\phi(e_1^*) \phi(e_2^*),\kappa'(e^*)\kappa(e^*)}, &\text{if}\; \kappa'(e) = \kappa(e) \;\text{for all}
\; e \neq e^*,\\
0, &\text{otherwise}.
\end{cases} \label{eq:tck}
\end{equation}

\uhead{Step 6.}
The prediction and update steps of the Bayesian filter are now
determined by the matrices $\tmp'$ and $\tmk'$.  For the prediction
step, suppose that the graph data through time $t_0$ is
$\kappa_{[0,t_0]}$ and let $\kappa = \kappa_{t_0}$ be a concise
notation for the graph at time $t_0$.  From a previous step of the
filter (or from an initialization) we are given the distribution on
the community assignments $\pr(\Phi_{t_0} | \Kappa_{[0,t_0]} =
\kappa_{[0,t_0]})$.  Starting with this distribution on $\phi$ at time
$t_0$, let $\pvkp_\kappa(t) \in \mathbb{R}^{m^n}$ (for all $t \ge
t_0$) be a vector whose $\phi$ component is the probability that (a)
the graph remains $\kappa$ during the time interval $[t_0,t)$, and (b)
the community assignment is $\phi$ at time $t$.  The initial value of
$\pvkp_\kappa$ is then $\pvkp_\kappa(t_0) = \pr(\Phi_{t_0} |
\Kappa_{[0,t_0]} = \kappa_{[0,t_0]})$.  Its evolution law is given by
\begin{equation}
\frac{d\pvkp_\kappa}{dt} = \tmp'_\kappa \pvkp_\kappa. \label{eq:predictDif}
\end{equation}
Note that $\tmp'_\kappa$ is not a transition rate matrix:  it allows
probability to leak out of the vector $\pvkp_\kappa(t)$ so that
its sum does not remain 1, but rather equals the probability
$\pr(\Kappa_{[t_0,t)} = \kappa | \Kappa_{[0,t_0]} =
\kappa_{[0,t_0]})$.  Normalizing $\pvkp_\kappa(t)$, however, gives us the
probability distribution of $\phi$ given that the graph has remained
$\kappa$ during the time interval $[t_0,t)$:
\begin{equation}
\pr\big(\Phi_t = \phi | \Kappa_{[0,t)} = \kappa_{[0,t)} \big) \propto
\big( \pvkp_\kappa(t) \big)_\phi. \label{eq:predict}
\end{equation}
This, then, is the prediction step of the Bayesian filter.  The update
step is obtained from $\tmk'$: given that the community assignment is
$\phi$, the probability of a single edge $e = \{v,w\}$ transitioning
from type $k = \kappa_{t^-}(e)$ to type $k' = \kappa_t(e)$ in some
infinitesimal time period $\Delta t$ is given by~\eqref{eq:tck} as
$b_{\phi(v) \phi(w),k'k} \Delta t$.  The conditional probabilities of
the posterior distribution on $\phi$ given this single edge transition
are proportional to this.  Therefore
\begin{equation}
\pr\big(\Phi_t = \phi | \Kappa_{[0,t]} = \kappa_{[0,t]} \big) \propto
b_{\phi(v) \phi(w),k'k} \pr\big(\Phi_{t^-} = \phi |
\Kappa_{[0,t)} = \kappa_{[0,t)} \big). \label{eq:update}
\end{equation}
This equation holds only for a single edge transitioning.  When
multiple edges transition at exactly the same time, the correct update
procedure would be to average, over all possible orderings of the edge
transitions, the result of applying~\eqref{eq:update} successively to
each transition.

\begin{figure}
  \centering
    \centerline{\includegraphics[width=0.95\columnwidth]{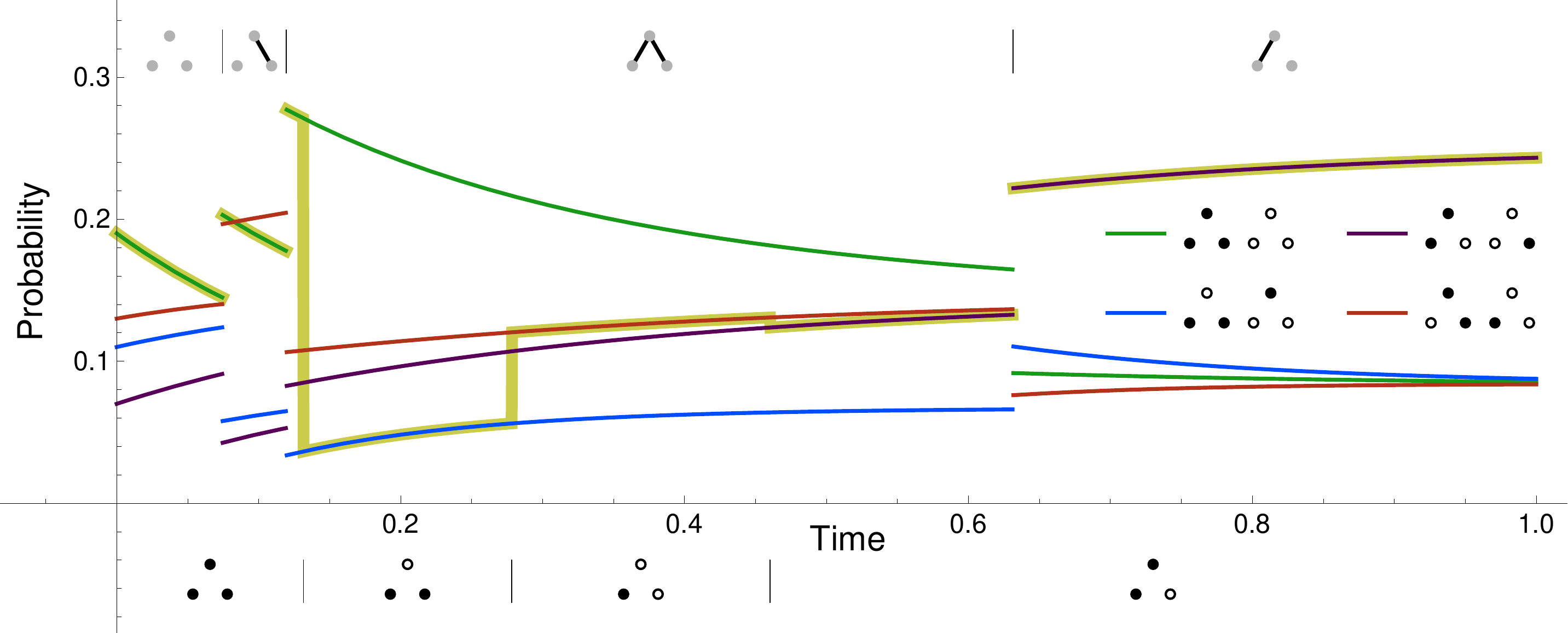}}
%
\caption{Temporal evolution of the probabilities of all community
  assignments $\phi$ for $\cH(3,2,1,3,1,1,3)$}
\label{fig:exactExample}
\end{figure}

Figure~\ref{fig:exactExample} shows the exact evolution for the very
simple, dynamic planted partition case $\cH(3,2,1,3,1,1,3)$.  Here
there are $n=3$ nodes and $m=2$ communities, so only four possible
partitions of the nodes (all in one community, or one of the three
nodes by itself).  Each partition corresponds to two community
assignments $\phi$ (of equal probability), and we plot the probability
$\pr(\Phi_t = \phi | \Kappa_{[0,t]} = \kappa_{[0,t]})$ for assignments
from each of these four partitions.  The graph data is shown in top
row of the figure: the graph is initially empty, then an edge turns
on, then another, and then an edge turns off.  We observe that while
the graph is some constant $\kappa$ the probabilities decay
exponentially toward a steady state vector (the null vector of
$\tmp'_\kappa$).  When the graph changes, the probability of each
community assignment hypothesis $\phi$ jumps, then begins decaying
toward a new steady state.  The bottom row of
Figure~\ref{fig:exactExample} shows the ground truth time-history
$\phi_t$ of community assignments which were used to generate the
graph data, and the yellow highlighting in the figure indicates which
community assignment hypothesis is the true one.  Further details may
be found in~\cite{Fer09}.

\subsection{Marginalization}
\label{sec:marginalization}

Using notation similar to that in Section~\ref{sec:staticApproxBlock},
let $p_i^v(t) \doteq \pr(\Phi_t(v) = i | \Kappa_{[0,t]} = \kappa_{[0,t]})$
be the probability that node $v$ is in community $i$ at time $t$,
$p_{ij}^{vw}(t) \doteq \pr(\Phi_t(v) = i, \Phi_t(w) = j | \Kappa_{[0,t]} =
\kappa_{[0,t]})$, and so on.  Note that the same notation $p_i^v(t)$
was used in Step 1 of the derivation in the previous section to denote
a prior probability, but henceforth it will indicate a quantity
conditioned on the graph data $\Kappa[0,t]$.  To indicate conditioning
on $\Kappa[0,t)$ we will use the notation $t^-$:  e.g.,
$p_i^v(t^-) \doteq \pr(\Phi_t(v) = i | \Kappa_{[0,t)} =
\kappa_{[0,t)})$.  The prediction and update equations for
the full probability distribution are linear (up to a constant
factor), so we can sum them over the groups of all nodes aside from
$v$ to obtain an expression for $p_i^v(t)$.  When we apply this
marginalization to~\eqref{eq:predictDif}, we get a quantity
$\tilde{p}_i^v(t)$ proportional to $p_i^v(t)$ (note that this use of
the notation $\tilde{p}_i^v$ differs from that of
Section~\ref{sec:staticApproxBlock}).  We let $\kappa \doteq \kappa_t$ be a
concise notation for the graph at time $t$.  The marginalization
of~\eqref{eq:predictDif} is then
\begin{equation}
\dot{\tilde{p}}_i^v =
\sum_{\iota=1}^m a_{i\iota} \tilde{p}_{\iota}^v +
\sum_{w \in [n] \atop w \neq v} \sum_{j=1}^m b_{ij,\kappa(\{v,w\})\kappa(\{v,w\})} \tilde{p}_{ij}^{vw} + \!\!\!
\sum_{\{w,x\} \subseteq [n] \atop w,x \neq v} \sum_{j=1}^m \sum_{k=1}^m b_{jk,\kappa(\{w,x\})\kappa(\{w,x\})} \tilde{p}_{ijk}^{vwx}.
\end{equation}
We may convert this to an equation for
$p_i^v(t)$ itself (albeit a nonlinear one) by expressing $p_i^v$ as
$p_i^v(t) = \tilde{p}_i^v(t) / P_\kappa(t)$.  The sum of
$p_i^v(t)$ from $i = 1$ to $m$ equals 1 for every node $v$, so the sum of
$\tilde{p}_i^v(t)$ from $i = 1$ to $m$ is $P_\kappa(t)$ for every $v$.
Because $\dot{p}_i^v(t) = \dot{\tilde{p}}_i^v(t) / P_\kappa(t) -
p_i^v(t) S_\kappa(t)$, where $S_\kappa(t) \doteq
\dot{P}_\kappa(t)/P_\kappa(t)$, we have
\begin{equation}
\begin{split}
\dot{p}_i^v =
&-p_i^v S_\kappa + \\
& \sum_{\iota=1}^m a_{i\iota} p_{\iota}^v +
\sum_{w \in [n] \atop w \neq v} \sum_{j=1}^m b_{ij,\kappa(\{v,w\})\kappa(\{v,w\})} p_{ij}^{vw} + \!\!\!
\sum_{\{w,x\} \subseteq [n] \atop w,x \neq v} \sum_{j=1}^m
\sum_{k=1}^m b_{jk,\kappa(\{w,x\})\kappa(\{w,x\})} p_{ijk}^{vwx}.
\end{split} \label{eq:predict1}
\end{equation}
Here $S_\kappa$ may be expressed as
\begin{equation}
S_\kappa(t) = \!\!\!\sum_{\{v,w\} \subseteq[n]} \sum_{i=1}^m \sum_{j=1}^m b_{ij,\kappa(\{v,w\})\kappa(\{v,w\})} p_{ij}^{vw}(t).
\end{equation}

We now marginalize the update equation~\eqref{eq:update} when the edge
$e$ transitions from type $l = \kappa_{t^-}(e)$ to type $l' =
\kappa_t(e)$.  To get the updated probability $p_i^v(t)$ that node $v$
is in community $i$ after the transition there are two cases: $v \in
e$ (say, $e = \{v,w\}$) and $v \notin e$ (say $e = \{w,x\}$):
\begin{equation}
p_i^v(t) = \frac{1}{S_{l'l}(t^-)} \begin{cases} \displaystyle
\sum_{j=1}^m b_{ij,l'l} p_{ij}^{vw}(t^-) & \text{if} \;\; e = \{v,w\}, \\ \displaystyle
\sum_{j=1}^m \sum_{k=1}^m b_{jk,l'l} p_{ijk}^{vwx}(t^-) \quad & \text{if} \;\; e = \{w,x\}.
\end{cases} \label{eq:update1}
\end{equation}
The normalization constant $S_{l'l}(t^-)$ is independent of the node $v$:
\begin{equation}
S_{l'l}(t^-) = \sum_{i=1}^m \sum_{j=1}^m b_{ij,l'l} p_{ij}^{e_1e_2}(t^-).
\end{equation}

Whereas the Bayesian filter~\eqref{eq:predict} and~\eqref{eq:update}
specifies the exact evolution of probabilities in an unmanageably
large state space ($m^n$ elements), the marginalized
filter~\eqref{eq:predict1} and~\eqref{eq:update1} involves only $mn$
elements.  It is still an exact filter---no approximations have been
made---but it is not useful as it stands because it is not closed.
The equations for the first-order statistics $p_i^v$ involve second-
and third-order statistics $p_{ij}^{vw}$ and $p_{ijk}^{vwx}$.  One
could write down equations for these, but they would involve still
higher-order statistics, and so on.  Instead, a {\em closure model} is
needed for the second- and third-order statistics in terms the
$p_i^v$.  The topic of closures is discussed in
Section~\ref{sec:dynamicApprox}.

\begin{figure}
  \centering
    \centerline{\includegraphics[width=0.4\columnwidth]{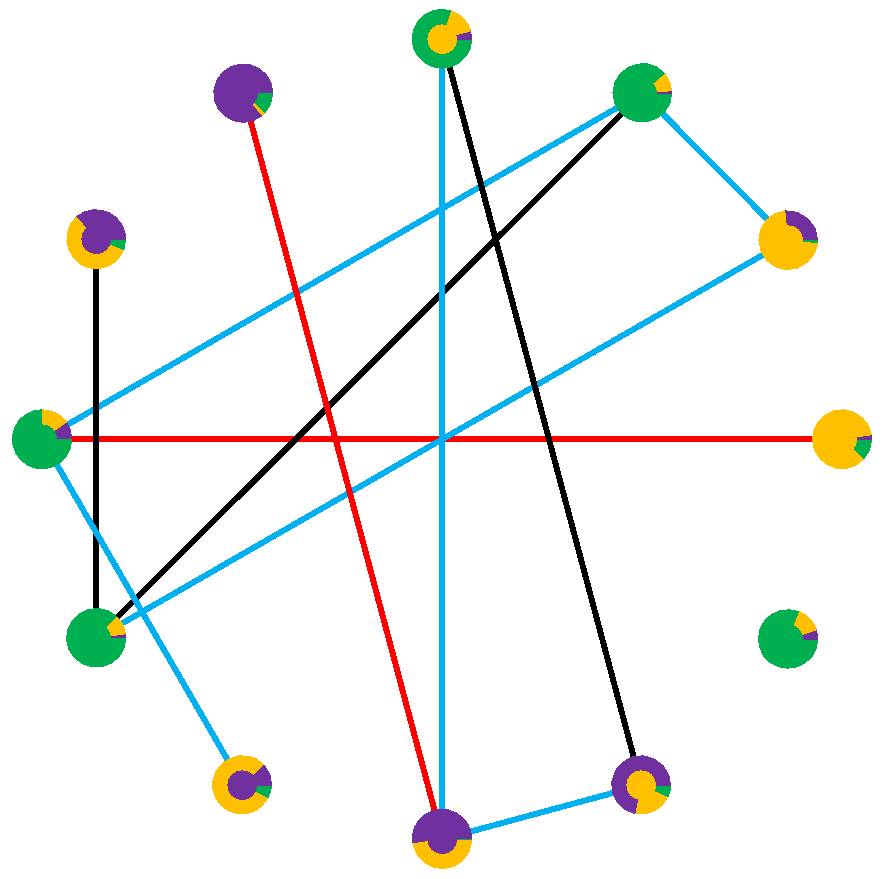}}
%
\caption{First-order marginals for community membership}
\label{fig:pies}
\end{figure}
To verify~\eqref{eq:predict1} and~\eqref{eq:update1}, one can run the
full filter~\eqref{eq:predict} and~\eqref{eq:update} and use it to
compute the required second- and third-order statistics.  The results
of evolving $p_i^v(t)$ using these oracular terms should then agree
with those obtained by marginalizing the full solution direction.
Figure~\ref{fig:pies} shows the results of such a verification.  A
simulation of $\cH(n,A,\bB)$ with $n=12$ nodes, $m=3$ communities, and
$r=4$ edge types was used to generate graph data $\kappa_{[0,0.8]}$
and ground-truth community assignment data $\phi_{[0,0.8]}$.  The
transition rate matrices $A$ and $B_{ij}$ used are given
in~\cite{Fer09}.  The final frame $(\phi_{0.8},\kappa_{0.8})$ of this
run is shown in Figure~\ref{fig:pies}: the centers of each disk
correspond to the communities (green, yellow, or purple) which
$\phi_{0.8}$ assigns to each node; the colors of each edge (white,
black, blue, or red) are given by $\kappa_{0.8}$.  The same parameters
$A$ and $\bB$ were then used in the Bayesian filter~\eqref{eq:predict}
and~\eqref{eq:update}, which required evolving a system of $3^{12}
\approx 530,000$ quantities.  The marginalized probabilities
$p_i^v(0.8)$ are depicted in the outer bands, so accuracy is indicated
by the outer band largely agreeing with the center.

In the case of the dynamic planted partition model
$\cH(n,m,a,\lambda_I,\mu_I,\lambda_O,\mu_O)$, the first-order
statistics are trivial: the prediction and update equations reduce to
the observation that the probability a node $v$ is in some community
equals 1.  Instead, with some bookkeeping, one can derive a Bayesian
filter for the second-order statistics $p_{ij}^{vw}(t)$ and reduce
these to a filter for the co-membership probabilities $\pvw(t)$: this
is similar to what was done in Section~\ref{sec:staticApproxPlanted},
although that was for an approximation based on limited graph
evidence, and this is exact.  The filter for $\pvw$ depends on third-
and fourth-order statistics.  There are 5 third-order statistics,
which sum to 1, and we denote them $p^{\{v\}\{w\}\{x\}}$,
$p^{\{v\}\{w,x\}}$, $p^{\{w\}\{v,x\}}$, $p^{\{x\}\{v,w\}}$, and
$p^{\{v,w,x\}}$.  These correspond to the probabilities that $v$, $w$,
and $x$ are in different communities, that two are in the same
community with $v$, $w$, and $x$, respectively, being in another, and
that all three are in the same community.  Similarly there are 15
fourth-order statistics.  The two that matter here are $p^{\{v,w,x,y\}}$
and $p^{\{v,w\}\{x,y\}}$.  The sum of these two is the probability that
$v$ is in the same community as $w$ and that $x$ is in the same
community as $y$.

The prediction step of the Bayesian filter for $\pvw(t)$ may be expressed as
\begin{equation}
\dot{p}^{\{v,w\}} = \frac{2am}{m-1}\left(\frac{1}{m} - \pvw\right)
- \gamma^{vw} q_{vw}^{vw} -
\overbrace{\sum_{x \in [n] \atop x \neq v,w}\!\big(\gamma^{vx} r_{vw}^{vx} + \gamma^{wx} r_{vw}^{wx} \big)}^{\displaystyle R_{vw}} -
\,\overbrace{ \!\!\!\sum_{\{x,y\} \subseteq [n] \atop x,y \neq v,w} \!\!\!\gamma^{xy} s_{vw}^{xy}}^{\displaystyle S_{vw}}. \label{eq:predict2}
\end{equation}
The form of the update step for $\pvw$ depends on whether the edge $e$
that is flipping at time $t$ has 2, 1, or 0 nodes in common with $\{v,w\}$:
\begin{equation}
\pvw(t) = \pvw(t^-) + \frac{1}{\delta^{e_1e_2}(t^-)} \begin{cases}
\gamma_{t^-}^{vw} q_{vw}^{vw}(t^-) & \text{if} \; e = \{v,w\}, \\
\gamma_{t^-}^{wx} r_{vw}^{wx}(t^-) & \text{if} \; e = \{w,x\}, \\
\gamma_{t^-}^{xy} s_{vw}^{xy}(t^-) & \text{if} \; e = \{x,y\}.
\end{cases} \label{eq:update2}
\end{equation}
The notation used in~\eqref{eq:predict2} and~\eqref{eq:update2} is
defined as follows.  We define $\gamma_{I,t}^{vw}$ to be the
transition rate for an edge to flip (i.e., turn on or off) between
nodes $v$ and $w$ at time $t$ under the hypothesis that they are in
the same community.  This transition rate depends on whether there is
currently an edge between $v$ and $w$.  Therefore,
$\gamma_{I,t}^{vw}$, and its counterpart $\gamma_{O,t}^{vw}$ for the
hypothesis that $v$ and $w$ are in different communities, are given by
\begin{equation}
\gamma_{I,t}^{vw} \doteq \begin{cases}
\mu_I & \text{if} \; \{v,w\} \in E(G_t),\\
\lambda_I & \text{if} \; \{v,w\} \in E(G_t),
\end{cases} \quad \text{and} \quad
\gamma_{O,t}^{vw} \doteq \begin{cases}
\mu_O & \text{if} \; \{v,w\} \in E(G_t),\\
\lambda_O & \text{if} \; \{v,w\} \in E(G_t).
\end{cases}
\end{equation}
The quantity $\gamma_t^{vw}$, which plays an important role
in~\eqref{eq:predict2} and~\eqref{eq:update2}, is the difference
between the flip rates under the two hypotheses:
\begin{equation}
\gamma_t^{vw} \doteq \gamma_{I,t}^{vw} - \gamma_{O,t}^{vw}.
\end{equation}
On the other hand, the normalization constant $\delta^{vw}(t)$
in~\eqref{eq:update2} is the expected flip probability given our
current knowledge of the probabilities of the two hypotheses:
\begin{equation}
\delta^{vw}(t) \doteq \gamma_{I,t}^{vw} \pvw(t) + \gamma_{O,t}^{vw} p^{\{v\}\{w\}}(t),
\end{equation}
where $p^{\{v\}\{w\}} = 1-\pvw$ is the probability of $v$ and $w$
being in different communities.  Finally, the $q$, $r$, and $s$
quantities represent modified second-, third-, and fourth-order
statistics, respectively:
\begin{align}
q_{vw}^{vw} &\doteq \pvw p^{\{v\}\{w\}},\\
r_{vw}^{vx} &\doteq p^{\{v,w,x\}} - \pvw p^{\{v,x\}}, \;\; \text{and}\\
s_{vw}^{xy} &\doteq p^{\{v,w,x,y\}} + p^{\{v,w\}\{x,y\}} - p^{\{v,w\}} p^{\{x,y\}}.
\end{align}
The $r$ and $s$ quantities measure deviations from independence.  That
is, if the event $\Phi_t(v) = \Phi_t(w)$ (i.e., $v$ and $w$ are in the
same group at time $t$) were independent of $\Phi_t(v) = \Phi_t(x)$,
then the probability of both events occurring (i.e., $\Phi_t(v) =
\Phi_t(w) = \Phi_t(x)$) would equal the product of their
probabilities: that is, $p^{\{v,w,x\}} = \pvw p^{\{v,x\}}$, or
$r_{vw}^{vx} = 0$.  Similarly, if $v$ and $w$ being in the same
community were independent of $x$ and $y$ being in the same community
(which seems more plausible), then we would have $s_{vw}^{xy} = 0$.
The terms $R_{vw}$ and $S_{vw}$ represent the accumulated effects on
$\dot{p}^{\{v,w\}}$ of the third- and fourth-order deviations from
independence, respectively.  Understanding the role of these
quantities is important for developing an effective closure for this
marginalized filter.

\section{Community Tracking:  Approximation}
\label{sec:dynamicApprox}

\begin{figure}
  \centering
    \centerline{\includegraphics[width=\columnwidth]{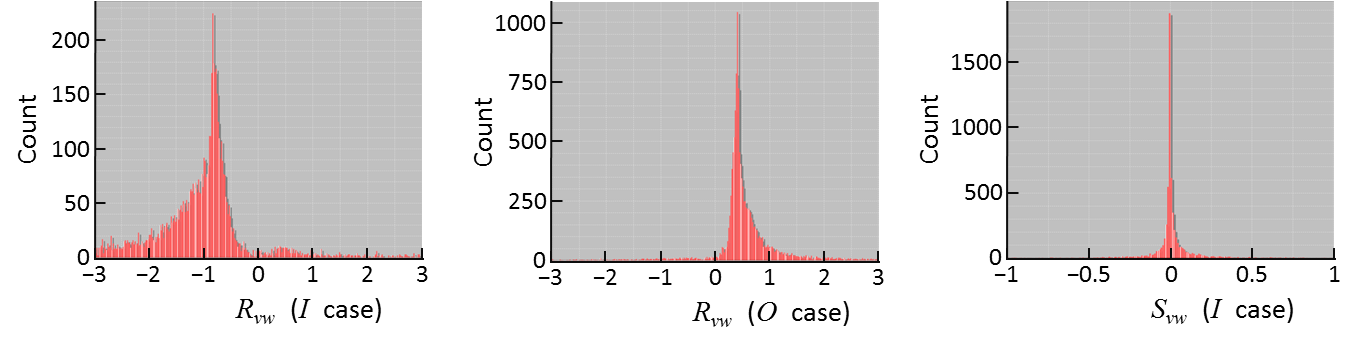}}
%
\caption{Histograms of third- and fourth-order statistics}
\label{fig:closures1}
\end{figure}

To develop closures for the $r_{vw}^{vx}$ and $s_{vw}^{xy}$ terms
in~\eqref{eq:predict2} and~\eqref{eq:update2} it helpful to know how
they behave statistically.  In this section we present a preliminary
investigation of these statistics and a possible closure for them
using, as an example, the $\cH(12,3,0.5,16,4,2,18)$ case depicted in
Figure~\ref{fig:rocket}.  For this case we compute histograms for the
$R_{vw}$ and $S_{vw}$ terms in~\eqref{eq:predict2}.  In each case, we
compile separate histograms for the case of $v$ and $w$ being in the
same ground-truth community (case $I$) and in different communities
(case $O$).  The histograms for the $I$ and $O$ cases of $R_{vw}$ are
shown in Figure~\ref{fig:closures1}.  In the $I$ case, we tend to have
$R_{vw} < 0$, which makes a positive contribution to
$\dot{p}^{\{v,w\}}$ in~\eqref{eq:predict2}; whereas in the $O$ case
the opposite occurs.  Thus these third-order statistics make an
important contribution to the evolution of $\pvw$.  On the other hand,
$S_{vw}$ appears to be much less important.  In the $I$ case, also
shown in Figure~\ref{fig:closures1}, $S_{vw}$ is tightly and
symmetrically clustered near 0.  The $O$ case is similar.  This makes
sense intuitively:  as mentioned at the end of
Section~\ref{sec:marginalization}, it seems more plausible for $v$ and
$w$ being in the same community to be independent of $x$ and $y$ being
in the same community than for $v$ and $w$ to be independent of $v$
and $x$.  Therefore, we will make this independence assumption to
obtain the fourth-order closure $s_{vw}^{xy} = 0$.  It remains to
develop a closure for $r_{vw}^{vx}$.

The five third-order statistics must sum to one and be consistent with
the second-order statistics.  This is expressed by the following four
equations:
\begin{align}
p^{\{v\}\{w\}\{x\}} + p^{\{v\}\{w,x\}} + p^{\{w\}\{v,x\}} +
p^{\{x\}\{v,w\}} + p^{\{v,w,x\}} &= 1, \label{eq:cons1} \\
p^{\{v\}\{w,x\}} + p^{\{v,w,x\}} &= p^{\{w,x\}}, \\
p^{\{w\}\{v,x\}} + p^{\{v,w,x\}} &= p^{\{v,x\}}, \;\; \text{and}\\
p^{\{x\}\{v,w\}} + p^{\{v,w,x\}} &= p^{\{v,w\}}. \label{eq:cons4}
\end{align}
This leaves one degree of freedom, which we choose $p^{\{v,w,x\}}$ to
represent.  The constraint that the variables
in~\eqref{eq:cons1}--\eqref{eq:cons4} are non-negative imposes the
following bounds on $p^{\{v,w,x\}}$:
\begin{align}
p^- &\doteq \frac{1}{2}\big(p^{\{w,x\}} + p^{\{v,x\}} + p^{\{v,w\}} \big),
\;\;\text{and}\\
p^+ &\doteq \min\big(p^{\{w,x\}}, p^{\{v,x\}}, p^{\{v,w\}} \big).
\end{align}
For consistency, a closure for $p^{\{v,w,x\}}$ should satisfy $p^- \le
p^{\{v,w,x\}} \le p^+$.  We will consider closure models that select
$p^{\{v,w,x\}} \in [p^-,p^+]$ as a function of $p^{\{w,x\}}$,
$p^{\{v,x\}}$, and $p^{\{v,w\}}$.  Many natural approximations (such
as a symmetric version of $p^{\{v,w,x\}} \approx p^{\{v,w\}}
p^{\{v,x\}}$) fail to satisfy $p^- \le p^{\{v,w,x\}} \le p^+$.  The
approximations $p^{\{v,w,x\}} \approx p^+$ and $p^{\{v,w,x\}} \approx
p_0^- \doteq \max(p^-,0)$ have poor properties.  A least-squares
solution is possible, but a better principle to employ is maximum
entropy~\cite{Jay03}.

When applying the maximum entropy principle, one needs a suitable
underlying measure space.  In this discrete case, this simply means a
set of atomic events which are equally likely {\it a priori}.  Such
events arise naturally in this case:  there are $m^3$ of them with
probabilities equal to $p_{ijk}^{vwx}$.  In the absence of graph data,
symmetry implies that their probabilities are each $m^{-3}$.  Thus the
entropy is defined as
\begin{equation}
H = -\sum_{i=1}^m \sum_{j=1}^m \sum_{k=1}^m p_{ijk}^{vwx} \log
p_{ijk}^{vwx}. \label{eq:H1}
\end{equation}
In this symmetric case, the $m^3$ values of $p_{ijk}^{vwx}$ take only
five distinct values, so we may re-write~\eqref{eq:H1} as
\begin{equation}
\begin{split}
H = &-p^{\{v\}\{w\}\{x\}} \log\frac{p^{\{v\}\{w\}\{x\}}}{(m)_3} -
p^{\{v\}\{w,x\}} \log\frac{p^{\{v\}\{w,x\}}}{(m)_2} -
p^{\{w\}\{v,x\}} \log\frac{p^{\{w\}\{v,x\}}}{(m)_2}\, + \\
&-p^{\{x\}\{v,w\}} \log\frac{p^{\{x\}\{v,w\}}}{(m)_2} -
p^{\{v,w,x\}} \log\frac{p^{\{v,w,x\}}}{m},
\end{split}
\end{equation}
where $(m)_r \doteq m!/(m-r)!$ denotes the falling factorial.  We may
use~\eqref{eq:cons1}--\eqref{eq:cons4} to express the other variables
in terms of $p^{\{v,w,x\}}$, then take the derivative of $H$ with
respect to $p^{\{v,w,x\}}$.  This reduces to
\begin{equation}
\frac{dH}{dp^{\{v,w,x\}}} = \log\left(\frac{(m-2)^2}{4(m-1)}
\frac{\big(p^{\{w,x\}} - p^{\{v,w,x\}}\big)\big(p^{\{v,x\}} -
  p^{\{v,w,x\}}\big)\big(p^{\{v,w\}} -
  p^{\{v,w,x\}}\big)}{p^{\{v,w,x\}} \big(p^{\{v,w,x\}}-p^-\big)^2}
\right).\label{eq:Hderv}
\end{equation}
Provided $p_0^- < p^+$, the function in~\eqref{eq:Hderv} is strictly
decreasing from $\infty$ to $-\infty$ on $[p_0^-,p^+]$, so it has a
unique zero within this interval, and this zero is where $H$ attains
its maximal value on $[p_0^-,p^+]$.  Finding this zero involves
solving a cubic equation, which yields the maximum entropy closure for
$r_{vw}^{vx}$.

\begin{figure}
  \centering
    \centerline{\includegraphics[width=0.8\columnwidth]{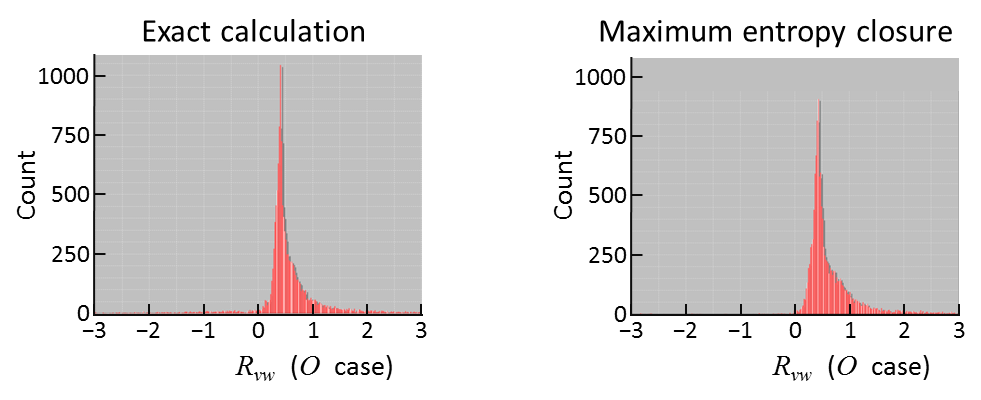}}
%
\caption{Third-order statistics:  exact vs. closure}
\label{fig:closures2}
\end{figure}

Figure~\ref{fig:closures2} indicates how well this closure
performed in the $\cH(12,3,0.5,16,4,2,18)$ example.  The histograms
show good agreement:  each has values that are
predominantly positive, with a peak in the same place, and
distributions of similar shapes, although the closure distribution is a
little more spread out.  This suggests that using this closure for
$r_{vw}^{vx}$ along with the fourth-order closure $s_{vw}^{xy}$ is a
promising idea.  Unfortunately, this closure has nothing built into it
that ensures $\pvw$ stays in the range $[0,1]$, and, indeed numerical
simulations that use it quickly produce probabilities outside this
range.

To see why it fails, consider the consistency requirement $p^- \le p^+$
mentioned above.  This may be re-written
\begin{equation}
\big| p^{\{v\}\{x\}} - p^{\{w\}\{x\}} \big| \le p^{\{v\}\{w\}} \le p^{\{v\}\{x\}} + p^{\{w\}\{x\}},
\end{equation}
i.e., the probabilities $p^{\{v\}\{w\}}$ satisfy the triangle
inequality.  In particular, if $\pvw = 1$, then $p^{\{v,x\}} =
p^{\{w,x\}}$ for all nodes $x \neq v,w$.  This makes sense: if $v$ and
$w$ are definitely in the same community $C$, then both $p^{\{v,x\}}$
and $p^{\{w,x\}}$ mean ``the probability that $x \in C$,'' so it would
be illogical for these values to differ.  Furthermore, in the special
case $a = 0$, if $\pvw = 1$ at some time, then it will remain 1, which
implies $\dot{p}^{\{v,x\}} = \dot{p}^{\{w,x\}}$.  The direct
verification of this fact using~\eqref{eq:predict2} leads to an
expression in which $S_{vw} = -R_{vw} \neq 0$.  In this case, the
fourth-order terms play a crucial role in maintaining consistency, so
it is not surprising truncating them entirely leads to
inconsistencies.  This is but one of the issues that must be addressed
before principled community tracking algorithms along these lines can
be developed, but we believe there is much promise in this approach of
Bayesian filtering using formal evolution and measurement models

\section{Conclusion}

Network science has benefited from the perspectives and expertise of
a variety of scientific communities.  The data fusion approach has
much to offer as well.  It provides an integrated framework for
synthesizing high-level situational awareness from messy, real-world
data.  It also develops tracking algorithms based on formal evolution
models that maintain representations of the uncertainty of the
ground-truth state.  We have applied this perspective to the community
detection problem in network science.  We began with a derivation of
the posterior probability distribution of community structure given
some graph data:  this is similar to the approach
of~\cite{Has06,HoWi08}.  However, rather than seeking the community
structure that maximizes this posterior probability, we developed
approximations to the marginals of these distributions.  In
particular, we consider the pairwise co-membership probability $\pvw$:
the probability that nodes $v$ and $w$ are in the same community.  We
develop an estimate of $\pvw$ based on using limited information from
the graph $G$ and approximating an integral over the model's
structural parameters.  The resulting method is very fast, and, when
exploited to produce a single community detection result (via the
utility formulation in Appendix~\ref{sec:utility}) yields
state-of-the-art accuracy.  Various uses for these $\pvw$ quantities
are combined in the network analysis and visualization product IGNITE.

We extended our community detection approach to tracking the
evolution of time-varying communities in time-varying graph data.  We
proposed dynamic analogs of the stochastic blockmodel and planted
partition models used in static community detection:  these models are
continuous-time Markov processes for the joint evolution of the
communities and the graph.  We derived a Bayesian filter for the current
probability distribution over all community structures given the previous
history of the graph data.  This filter decomposes into prediction
steps (during periods of constant graph data) and update steps (at
times when the graph changes).  The filter is over too large a state
space to use directly, so we marginalized it to get state spaces of a
reasonable size.  These marginalized equations require closures for
their higher-order terms, and we discussed one possible closure based
on maximum entropy.

The community detection work could be extended to more realistic graph
models, such as the degree-corrected blockmodel of~\cite{KaNe11}, and
the integral approximation developed in
Section~\ref{sec:staticApproxPlanted} could certainly be improved.
There is much more work to do in the community tracking case, however.
In the spectrum of methods that handle dynamic network data, the
models presented are intermediate between those that use the data only
to discern a static community structure (e.g.,~\cite{SCK07}) and those
mentioned in Section~\ref{sec:dynamicExact} that allow not just
individual node movements, but the birth, death, splitting, and
merging of communities.  Extending this methodology to account for
these phenomena would be important for applications.  Whatever
model is used, some method for parameter selection must be developed.
Integrating over the parameter space may be too complicated in the
community tracking case, but one may be able to extend the Bayesian
filter by making the input parameters themselves hidden variables, and
use Hidden Markov Model (HMM) techniques for parameter
learning~\cite{BaPe66}.  Closures must be developed,
preferably with proofs of their properties, rather than just
experimental justification.  Finally, to be truly useful, community
tracking need to be able to incorporate ancillary, non-network
information about the properties of the nodes and edges involved.
Bayesian data fusion provides an excellent framework for coping with
this kind of practical problem by identifying what expert knowledge is
needed to extend the uncertainty management in a principled manner.

\section*{Acknowledgments}
Research partially supported by ONR Contract N0001409C0563.  The
material presented here draws on the conference
papers~\cite{Fer09,FeBu10,FBA11}.  The authors thank Andrea
Lancichinetti for the LFR benchmark data shown in
Figure~\ref{fig:GFcompare}.

{\small
\providecommand{\bysame}{\leavevmode\hbox to3em{\hrulefill}\thinspace}
\providecommand{\MR}{\relax\ifhmode\unskip\space\fi MR }
\providecommand{\MRhref}[2]{%
  \href{http://www.ams.org/mathscinet-getitem?mr=#1}{#2}
}
\providecommand{\href}[2]{#2}

}

\appendix

\section{Expected Utility Formulation}
\label{sec:utility}

\begin{figure}[b]
  \centering
    \centerline{\includegraphics[width=0.45\columnwidth]{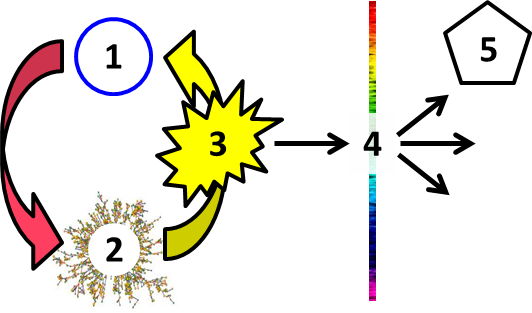}}
%
\caption{Five-part inference framework}
\label{fig:fiveStep}
\end{figure}

Figure~\ref{fig:fiveStep} depicts the data fusion approach this paper
takes toward community detection and tracking.  Part (1) represents
the model of the hidden state space in which we are interested (in
this case the community assignments $\phi$ or community partitions
$\pi$), including some prior probability distribution over it, and, in
the dynamic case, an evolution model.  Part (2) represents observed
data, and the red arrow from (1) to (2) is the measurement model which
assigns conditional probabilities of the data given the true state.
Part (3) is Bayesian inversion, which yields the conditional
probabilities of the true state given the data.  When the number of
possible states is large, the combinatorial explosion of part (3) must
be dealt with in some manner.  Thus, part (4) is some reduced
representation of the full posterior distribution in (3):  in this
case the pairwise co-membership probabilities $\pvw$.  Finally, part (5)
represents the end user of this inference process.  The full posterior
could, in principle, answer every question the end user might have
about the true state, given the limitation of the data available.  The
reduced representation (4) should be chosen in such a way that it may
be gleaned efficiently from (3), but also meet the needs of the end
user (5).

The body of this paper addresses parts (1)--(4) for the community
detection and tracking problems.  This appendix connects that analysis
to part (5).  Bayesian decision theory~\cite{Ber93} provides a way to
formalize the connection.  Consider the problem of how to assess the
quality of some proposed (or computed) partition $\hpi$.  The usual
approach is to stipulate some quality function of $\hpi$ and the graph
$G$.  Bayesian decision theory argues for assessing the quality of
$\hpi$ in terms of its agreement with the true partition $\pi$.  Thus,
with the cooperation of the end user in (5), we should construct a
{\em utility function} $U(\hpi|\pi)$ which defines the value of being
provided the answer $\hpi$ when the true partition is $\pi$.  Because
the ground truth partition $\pi$ is unknown, we cannot compute $U(\hpi|\pi)$.
However, if we know the posterior distribution $\pr(\pi|G)$, we can
compute the {\em expected utility} $\mathbb{E}[U](\hpi|G)$.  In
general, summing over all possible partitions $\pi$ to get this
expected utility is not practical, but we shall provide a generic
example in which the summation reduces to one over all pairs of
nodes, with the set of pairwise co-membership probabilities $\pvw$
providing all necessary statistical information.  In this sense, the
values of $\pvw$ comprise the reduced representation in part (4) of
Figure~\ref{fig:fiveStep}.

Let $V = [n]$ denote the set of $n$ nodes, and $\Pi(V)$ denote the set of
all partitions of $V$.  For any particular $\pi \in \Pi(V)$, let
$U(\hpi|\pi)$ denote the utility of the decision that $\hpi$ is the
correct partition when $\pi$ is in fact correct.  For any graph $G$ on
$V$, one may define the quality of $\hpi$ for $G$ to be the expected
value of its utility:
\begin{equation}
  \label{eq:eudef}
  \mathbb{E}[U](\hpi|G) = \!\! \sum_{\pi \in \Pi(V)} \!\! U(\hpi|\pi)
\pr(\pi | G),
\end{equation}
where $\pr(\pi|G)$ is the posterior probability of $\pi$ given $G$.
To simplify~\eqref{eq:eudef}, assume $U(\hpi|\pi)$ is additive: i.e.,
that it can be expressed as
\begin{equation}
  \label{eq:udef}
  U(\hpi|\pi) = \sum_{\hC \in \hpi} u(\pi[\hC])
\end{equation}
for some function $u$, where $\pi[\hC]$ is the partition of $\hC$
induced by $\pi$ (that is, the sets into which $\pi$ ``chops up''
$\hC$).  In this case, one can compute the expected utility of each
community $\hC \in \hpi$ given $G$,
\begin{equation}
  \label{eq:eucg}
  \mathbb{E}[U](\hC|G) = \!\! \sum_{\pi \in \Pi(V)} \!\! u(\pi[\hC])
\pr(\pi | G),
\end{equation}
and sum them over $\hC \in \hpi$ afterward to get
$\mathbb{E}[U](\hpi|G)$.  Grouping the sum in~\eqref{eq:eucg}
into classes based on the value of $\pi[\hC]$ yields
\begin{equation}
  \label{eq:eucgrho}
  \mathbb{E}[U](\hC|G) = \!\! \sum_{\rho \in \Pi(\hC)} \!\! u(\rho)
  \!\!\!\! \sum_{\pi : \pi[\hC] = \rho} \!\!\! \pr(\pi | G).
\end{equation}
The inner sum in~\eqref{eq:eucgrho} is the probability of $\rho$ given $G$,
denoted $\pr(\rho | G)$.  By grouping the sum in~\eqref{eq:eucg} less
aggressively, however, one obtains a more general result which
includes~\eqref{eq:eucg} and~\eqref{eq:eucgrho} as special cases.  For any
$\hC^+ \supseteq \hC$,
\begin{equation}
  \label{eq:eucgrhoplus}
  \mathbb{E}[U](\hC|G) = \!\!\!\! \sum_{\rho^+ \in \Pi(\hC^+)}
  \!\!\!\! u(\rho^+[\hC]) \pr(\rho^+ | G).
\end{equation}
In particular, we let $\hC^+ = \hC \cup \{v\}$ and define
\begin{equation}
  \label{eq:excvdef}
  \excv(\hC,v) = \mathbb{E}[U](\hC^+|G) - \mathbb{E}[U](\hC|G) - \mathbb{E}[U](\{v\}|G)
\end{equation}
to be the increase in expected utility of merging the community $\hC$
with the singleton community $\{v\}$.  Using~\eqref{eq:eucgrhoplus},
$\excv(\hC,v)$ may be expressed as
\begin{equation}
  \label{eq:excv}
   \excv(\hC,v) = \!\!\!\! \sum_{\rho^+ \in \Pi(\hC^+)} \!\!\!\! \left(
   u(\rho^+) - u(\rho^+[\hC]) - u(\{\{v\}\}) \right) \pr(\rho^+ | G).
\end{equation}
This equation allows the expected utility of a community to be computed by
adding one node at a time.

\subsection{A generic utility model}

Here we consider a simple, generic utility function in which some
node $v$ has been identified as ``bad,'' implying that the nodes
in his ground-truth community $C = \pi(v)$ are bad as well, but that
decision makers are using some other partition $\hpi$, resulting in all
nodes in $\hC = \hpi(v)$ ending up ``dead.''  Let
$u_{BD}$ and $u_{BA}$ be the respective utilities of a bad node
being dead and being alive, and $u_{GD}$ and $u_{GA}$ be the
corresponding utilities for good nodes.  Summing the utilities
over the cases $v \in V$ being identified as bad yields the utility
function
\begin{equation}
  \label{eq:ussimp}
  \begin{split}
  U^*(\hpi|\pi) = &(u_{BD} - u_{BA} - u_{GD} + u_{GA}) \!\!\! \sum_{C \in
    \pi, \hC \in \hpi} \!\!\!\! |C \cap \hC|^2 + \\
   &(u_{BA} - u_{GA}) \sum_{C \in \pi} |C|^2 +
   (u_{GD} - u_{GA}) \sum_{\hC \in \hpi} |\hC|^2 +
   u_{GA} n^2
  \end{split}
\end{equation}
after some rearrangement.  Assuming $u_{GA} \ge u_{GD}$ and $u_{BD}
\ge u_{BA}$ (and equality does not hold for both), a
convenient choice of translation and scaling yields
\begin{equation}
  \label{eq:usnondim}
  U(\hpi | \pi) = \frac{1}{2} \bigg( \;
    \sum_{C \in \pi, \hC \in \hpi} \!\!\!\! |C \cap \hC|^2 -
    \theta \sum_{\hC \in \hpi} |\hC|^2 - (1-\theta) n \bigg),
\end{equation}
where the threshold $\theta =
(u_{GA}-u_{GD})/(u_{GA}-u_{GD}+u_{BD}-u_{BA}) \in [0,1]$ reflects the
emphasis of killing bad nodes (small $\theta$, yielding large
communities) versus not killing good nodes (large $\theta$, small
communities).  This utility function is additive: it may be decomposed
according to~\eqref{eq:udef} with
\begin{equation}
  \label{eq:urho}
  u(\rho) = \frac{1}{2} \bigg(\;
    \sum_{C \in \rho} |C|^2 -
    \theta |\hC|^2 - (1-\theta) |\hC| \bigg),
\end{equation}
for any $\rho \in \Pi(\hC)$.  Note that $u(\rho) = 0$ for singleton
partitions $\rho = \{\{v\}\}$.

Substituting into~\eqref{eq:urho} into~\eqref{eq:excv} yields
\begin{equation}
  \label{eq:excv2}
   \excv(\hC,v) = \!\!\! \sum_{\rho^+ \in \Pi(\hC^+)} \!\!
   \Big(|\rho^+(v)| - 1 - \theta |\hC| \Big) \pr(\rho^+ | G).
\end{equation}
This may be written $\excv(\hC,v) = \mathbb{E}[\rho^+(v)-1 | G] -
\theta |\hC|$.  The first term here is the expected number of nodes in
$\hC$ that are in the same community as $v$ in a random partition of
$\hC^+$, given the evidence $G$.  Linearity of expectation implies
\begin{equation}
  \label{eq:exprhop}
  \mathbb{E}[\rho^+(v)-1 | G] = \sum_{w \in \hC} \pvw,
\end{equation}
where $\pvw = \pr(\{\{v,w\}\}|G)$ is the probability that $v$
and $w$ lie in the same community, given $G$.  This may be used to build
up $\mathbb{E}[U](\hC|G)$ by adding nodes $v$ to $\hC$ one at a time,
which yields the simple formula
\begin{equation}
  \label{eq:utilresult}
  \mathbb{E}[U](\hC|G) = \!\!\!\sum_{\{v,w\} \subseteq \hC}\!\!\!
  \big(\pvw - \theta \big).
\end{equation}
Summing this over $\hC$ yields an expression for the expected utility
of a partition $\hpi$ given the graph $G$ which involves only sums
over pairs of nodes in the same community $\hC \in \hpi$:
\begin{equation}
  \mathbb{E}[U](\hpi|G) = \!\sum_{\hC \in \hpi} \!\!\sum_{\{v,w\} \subseteq \hC}\!\!\!
  \big(\pvw - \theta \big). \label{eq:fullutilresult}
\end{equation}

\end{document}